\documentclass{aa} 
\usepackage[varg]{txfonts}
\usepackage{natbib}
\usepackage{xcolor}
\usepackage{subcaption}

\usepackage{graphicx}	
\usepackage{amsmath}	
\usepackage{amssymb}	
\usepackage{makecell}
\usepackage{placeins}
\usepackage{longtable}
\usepackage{lipsum}
\usepackage{txfonts}
\usepackage{threeparttable}
\usepackage[bottom]{footmisc}
\usepackage[labelfont=bf]{caption}
\begin{document} 

   \title{The VVV survey: Long-period variable stars}

   \subtitle{I. Photometric catalog of ten VVV/OGLE tiles}

   \author{F. Nikzat\inst{\ref{inst1},\ref{inst2}}, C. E. Ferreira Lopes\inst{\ref{inst3}}, M. Catelan\inst{\ref{inst1},\ref{inst2},\ref{inst4}}, R. Contreras Ramos\inst{\ref{inst1},\ref{inst2}}, M. Zoccali\inst{\ref{inst1},\ref{inst2}},\\ A. Rojas-Arriagada\inst{\ref{inst1},\ref{inst2}}, V. F. Braga\inst{\ref{inst5},\ref{inst6}}, D. Minniti \inst{\ref{inst7},\ref{inst7b}}, J. Borissova\inst{\ref{inst2},\ref{inst8}}, and I. Becker\inst{\ref{inst1},\ref{inst9}}
   }
   
   \institute{Instituto de Astrof\'isica, Facultad de F\'isica, Pontificia Universidad Cat\'olica de Chile, Av. Vicu\~na Mackenna~4860, 7820436 Macul, Santiago, Chile;  \email{fnikzat@astro.puc.cl}\label{inst1}
         \and
   Millennium Institute of Astrophysics, Nuncio Monse\~{n}or Sotero Sanz 100, Of. 104, Providencia, Santiago, Chile \label{inst2}
         \and
   Instituto Nacional de Pesquisas Espaciais (INPE/MCTIC), Av. dos Astronautas, 1758, São José dos Campos, 12227-010 SP, Brazil\label{inst3}
         \and
   Centro de Astroingenier{\'{\i}}a, Pontificia Universidad Cat{\'{o}}lica de Chile, Av. Vicu\~{n}a Mackenna 4860, 7820436 Macul, Santiago, Chile\label{inst4}
         \and
   INAF-Osservatorio Astronomico di Roma, via Frascati 33, 00078 Monte Porzio Catone, Italy\label{inst5}
         \and
   Space Science Data Center, via del Politecnico snc, 00133 Roma, Italy\label{inst6}
   \and
      Departamento de Ciencias Físicas, Facultad de Ciencias Exactas, Universidad Andr\'es Bello, Fern\'andez Concha 700, Las Condes, Santiago, Chile\label{inst7}
   \and
   Vatican Observatory, V00120 Vatican City State, Italy\label{inst7b}
   \and
      Instituto de F\'isica y Astronom\'ia, Universidad de Valpara\'iso, Av. Gran Breta\~{n}a 1111, Playa Ancha, Casilla 5030, Chile\label{inst8}
   \and
   Departamento de Ciencias de la Computaci\'on, Pontificia Universidad Cat\'olica de Chile, 7820436 Macul, Santiago, Chile\label{inst9}
            }
%


\authorrunning{F. Nikzat et al.}
\titlerunning{LPVs in the VVV survey}

  \abstract
   {Long-period variable stars (LPVs) are pulsating red giants, primarily in the asymptotic giant branch phase, and they include both Miras and semi-regular variables (SRVs). Their period-age and period-luminosity relations enable us to trace different stellar populations, as they are intrinsically very bright and cover a wide range in distances and ages.}
   {The purpose of this study is to establish a census of LPV stars in a region close to the Galactic center, using the six-year database of the Vista Variables in the V\'ia L\'actea (VVV) ESO Public Survey, as well as to describe the methodology that was employed to search for and characterize LPVs using VVV data. Near-IR surveys such as VVV provide a unique opportunity to probe the high-extinction innermost regions of the Milky Way. The detection and analysis of the intrinsically bright Miras in this region could provide us with an excellent probe of the properties of the Milky Way far behind its bulge.} 
   {We used point-spread function photometry for all available $K_{s}$-band images in ten VVV tiles, covering $16.4~\deg^2$ in total, overlapping fields observed in the course of the Optical Gravitational Lensing Experiment (OGLE)-III survey. We designed a method to select LPV candidates, and we used the known variables from OGLE-III and other known variables from the literature to test our approach. The reduced $\chi^2$ statistic, along with the flux-independent index $K_{(fi)}$, were used in our analysis. The Lomb-Scargle period search method, Fourier analysis, template fitting, and visual inspection were then performed to refine our sample and characterize the properties of the stars included in our catalog.}
   {A final sample of 130 Mira candidates, of which 129 are new discoveries, was thus obtained, with periods in the range between about 80 and 1400~days. Moreover, a sample of 1013 LPV candidates is also presented, whose periods are however not sufficiently constrained by the available data.  A fraction of the latter may eventually turn out to be SRVs. Ages are measured for these stars based on a reassessment of the period-age relations available in the literature. The Miras in our catalog include 18 stars satisfying the requirements to serve as reliable distance indicators and which are not saturated in the VVV $K_{s}$-band images. Their distances are accordingly derived and discussed. A number of objects that are seemingly placed far behind the Milky Way's bulge was detected.}
   {}


   \maketitle

\section{Introduction}\label{sec:intro}
In their late stages of evolution, prior to reaching the white dwarf cooling track, most low- and intermediate-mass stars go through the asymptotic giant branch (AGB) phase \citep[e.g.,][and references therein]{Iben-1983, Catelan-2015book}. When such stars undergo high-amplitude brightness variations, they are generally classified as long-period variables (LPVs), given the long periods, in excess of about 80~d, that are typically observed. More massive stars may also undergo LPV pulsations while on the red supergiant and/or super-AGB phase, and thus at least some LPVs may also be  core-collapse supernova progenitors \citep[][and references therein]{Levesque-2017,OGrady-2020}. LPV-like pulsations have also been observed in other, more exotic objects whose physical nature is still not fully understood \citep[][and references therein]{OGrady-2020}.

Based on their light-curve morphology, the range of their amplitudes and their periodicity, LPVs are commonly subgrouped into four categories: Miras, semi-regular variables (SRVs), irregulars (Irrs), and OSARGs (OGLE small-amplitude red giants). A recent review was provided by \citet{Catelan-2015book}, to which we refer the reader for further details and references. In our work, we mainly focus on Miras presenting periods of 80-1400~days and amplitudes greater than approximately 0.4~mag in the $K_s$-band. Their initial masses have a lower limit of about $0.8-1 \, M_\odot$ and an upper limit of $6-9 \, M_\odot$, with ages between $\sim 10^8$ and $\sim 10^{10}$~yr \citep[][]{Iben-1983, Catelan-2015book}.

Since \citet[][]{Glass-1981} and \citet[][]{Feast-1989} first presented the Miras period-luminosity relation, this class of variable stars has been used as distance indicators, and thus as tracers of the structure of the different Milky Way components \citep[e.g.,][]{Catchpole-2016, grady-2020, Urago-2020}.
Recently, {\em Gaia}'s Data Release 2 (DR2) provided a catalog of LPV candidates based on {\em Gaia} data collected over a time span of 22 months \citep[][]{Mowlavi-2018}. The {\em Gaia} DR2 LPV catalog includes both Miras and SRVs with variability amplitudes larger than 0.2~mag in the {\em Gaia} $G$-band. Accurate parallaxes are also included, which can be used to obtain their distances \citep[but see also][]{Whitelock-2012}.
On the other hand, Miras also display a period-age relation \citep[][]{Feast-2000,Feast-2007}, which, together with the fact that they span a broad range in ages and follow a period-luminosity relation, can make them useful probes not only of Galactic structure but also of the Galaxy's evolution. Since ages thus determined depend on the period, they are insensitive to systematic errors that may affect the reddening or distances. Even relative age estimates for an abundant variable population can supply constraints on the existence of age gradients. Using Miras as age indicators provides an opportunity not only of mapping the Galactic disk and bulge as a function of age but also the halo structure and even farther away, since Miras are intrinsically bright and detectable out to large distances.

Miras are classified into two subclasses based on their surface C/O abundance ratio: O-rich Miras, presenting H$_{2}$O, TiO, and SiO on their surfaces, and C-rich Miras, presenting C$_{2}$ and CN, with the C being dredged up from the stellar interior to the surface. In the Galactic bulge, the Mira population is dominated by the O-rich subclass, with only a few C-rich Miras being known \citep[][]{whitelock-2006,Matsunaga-2017,grady-2020}. In the lower-metallicity environments of the Magellanic Clouds, on the other hand, the proportion of C-rich Miras is larger \citep[e.g., ][]{whitelock-2003}.

It has long been known empirically that O-rich Mira kinematics is correlated with their periods in the Galactic disk. Thus, the relationship between age and velocity dispersion calibrated for stars in the solar neighborhood can be used to derive a period-age relationship for nearby O-rich Miras  \citep[][]{Feast-2000,Feast-2007}. Recently, \citet[][]{Grady-2019} introduced a period-age relation for Miras based on a tentative association between some of these stars and star clusters. Their resulting relation presented a large offset in comparison to other expressions available in the literature \citep[][]{Feast-2007,Lopez-Corredoira-2017} dealing with ages versus kinematics. That notwithstanding, all these works consistently find that longer-period Miras are younger, as expected since the period of a Mira increases with increasing luminosity, and hence mass.

   \begin{figure*} [t]
   \centering
   \includegraphics[width=0.975\hsize]{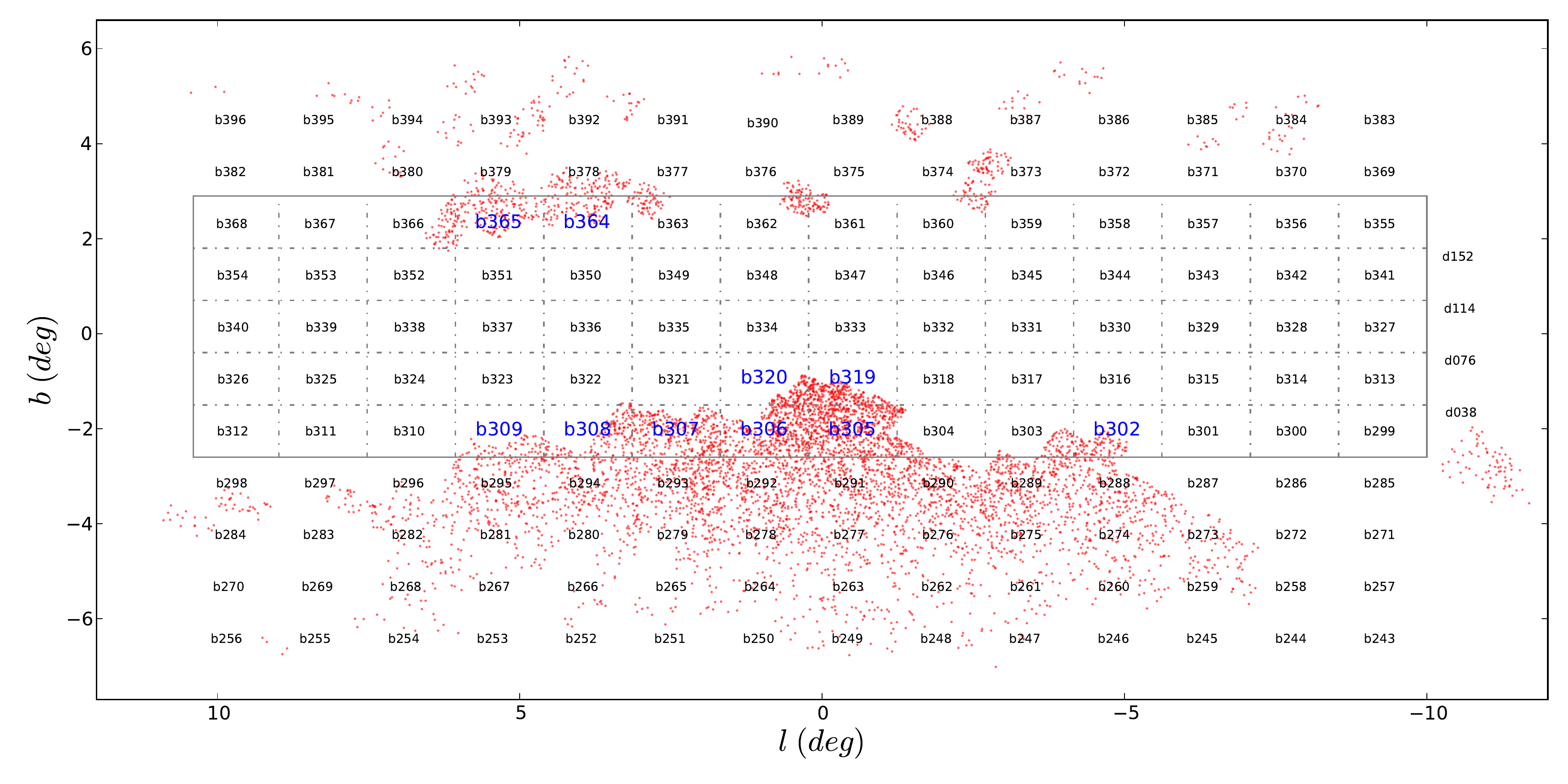}
   \caption{Footprint of the VVV survey in the central bulge of the MW, expressed in Galactic coordinates. Each VVV tile is marked by its respective ID in this diagram, those analyzed in this work being highlighted in blue and using a larger font size. Red dots represent stars classified as Miras in the OGLE-III project. The gray grids present a region of the VVV footprint where our PSF photometry is already completed.}
              \label{Fig:map}
    \end{figure*}

To study long-term variability phenomena in the bulge, a long-baseline near-infrared (IR) survey is desired. The IR regime allows one to overcome the high extinction levels along these sight lines, and the long baseline affords multiepoch photometry covering long variability timescales. Prior near- and mid-IR surveys, such as the Two-Mass All-Sky Survey  \citep[2MASS;][]{Skrutskie-2006}, 
DENIS \citep[Deep Near Infrared Survey of the Southern Sky; ][]{Epchtein-1999},
GLIMPSE \citep[Galactic Legacy Infrared Mid-Plane Survey Extraordinaire;][]{Benjamin-2005}, and UKIDSS-GPS \citep[United Kingdom Infrared telescope Deep Sky Galactic Plane Survey;][]{Lucas-2008}, could alleviate the problem of high extinction in the inner part of the Milky Way  \citep[MW; see also, e.g.,][]{Groenewegen-2005, J-Esteban-2015}, but still the absence of multiepoch observations hindered detailed analyses of variable stars, including LPVs. 
On the other hand, optical long-baseline surveys of the bulge region, such as OGLE \citep[Optical Gravitational Lensing Experiment;][]{Udalski-2015}, MACHO \citep[Massive Compact Halo Objects survey;][]{Alcock1996}, EROS \citep[Exp\'erience pour la Recherche d'Objets Sombres;][]{Aubourg-1993}, and {\em Gaia} \citep[Global Astrometric Interferometer for Astrophysics;][]{Mowlavi-2018}, provided a great demonstration of the power of multiepoch observations, with a plethora of important results for numerous different types of variable stars. On the other hand, the inner part of the bulge could not be penetrated due to the severe extinction in that region. Thus, even in the {\em Gaia} era, IR observations remain crucial to properly understand the stellar populations of the innermost MW. This is especially true in the case of faint stars or variables close to minimum light.

The VISTA Variables in the V\'ia L\'actea (VVV) ESO Public Survey \citep{Minniti-2010} was the first systematic, wide-field near-IR variability survey of the MW bulge and inner disk. A total of 1929 hours of observations were obtained between 2010-2016, yielding on average about 100 datapoints per field, and a total sky coverage of order $520 \, \deg^2$. Other near-IR variability studies of MW bulge fields, by comparison, have sometimes covered sky areas not exceeding $1 \, \deg^2$ \citep[e.g.,][]{Matsunaga-2009}. In addition to its extended sky coverage, the VVV images offer significantly improved resolution and higher depth by up to several magnitudes, as compared to 2MASS \citep[e.g.,][]{Saito-2012}.  

In this paper, we explore the VVV dataset in order to extract LPVs in regions presenting overlap with the OGLE footprint. We provide a first catalog of LPV candidates in this field, including periods and (when possible) distances. In future papers of this series, we will expand our search toward other regions covered by the VVV survey, with the ultimate goal of building a comprehensive catalog of LPVs covering the entire VVV footprint and studying their spatial distribution and ages.

\section{Observations and data reduction}\label{sec:obsdata}

Observations were taken with the 4\,m Visible and Infrared Survey Telescope for Astronomy (VISTA) at ESO’s Cerro Paranal Observatory, northern Chile, using the wide-field VISTA InfraRed Camera \citep[VIRCAM;][]{Dalton-2006, Emerson-2010} in the time span of 2010-2016. The data consist of multiepoch $K_s$ images and a couple of observations in each of the $ZYJH$ bandpasses per field. Each exposure (commonly referred to as a ``pawprint image'') is taken with a mosaic of 16 CCDs covering $2048 \times 2048$ pixels each at 0.34\arcsec\ resolution, with exposure time 5s. The seeing value is typically around 1\arcsec. Multiple pawprint exposures of a given field are later combined to generate so-called ``tile images,'' which (by design) remove the gaps that are present in between individual CCDs in the pawprint images. Each such tile covers $1.64 \, \deg^2$ \citep{Minniti-2010}. Disk tile names are labeled by a $d$ followed by an ID number ranging between 001 and 152; bulge tiles are instead denoted by a $b$, followed by an ID number in the range 201-396 \citep[e.g.,][]{Catelan-2011,Saito-2012}.

\begin{table}
\caption{VVV tiles analyzed in this paper \tablefootmark{(a)} }   
\label{table:coo}      
\centering  
\begin{tabular}{l r r r r}
\hline
\hline
  \multicolumn{1}{c}{Tile} &
  \multicolumn{1}{c}{$\ell$[deg]} &
  \multicolumn{1}{c}{$b$[deg]} &
  \multicolumn{1}{c}{$N_{\rm OGLE-Mira}$} &  
  \\
\hline
\hline
  b305 & -0.507 & -2.045 & 519 \\
  b306 &  0.952 & -2.045 & 336 \\
  b319 & -0.510 & -0.953 & 260 \\
  b307 &  2.412 & -2.045 & 255 \\
  b365 &  5.341 &  2.324 & 131 \\
  b308 &  3.871 & -2.044 & 125 \\
  b364 &  3.881 &  2.324 & 77 \\
  b302 & -4.885 & -2.045 & 76 \\
  b320 &  0.948 & -0.953 & 70 \\
  b309 &  5.330 & -2.045 & 58 \\
\hline
\hline
\end{tabular}
\tablefoot{
\tablefoottext{a}{
VVV designation and central Galactic coordinates $(\ell, b)$ of VVV tiles in which over 50 Miras detected by the OGLE-III survey are located. The number of OGLE Miras ($N_{\rm OGLE-Mira}$) in each tile is also presented.}}
\end{table}

Due to the very high stellar density in our studied fields, we consider point-spread function (PSF) photometry to be superior to aperture photometry, as adopted in some previous variability studies that also used VVV data \citep[e.g.,][]{Saito-2013, Teixeira-2018, Dekany-2019, FerreiraLopes-2020, Herpich-2021}. Accordingly, PSF photometry was obtained for all sources by fitting appropriate models on individual pawprint images, using the {\sc daophot/allstar} software package  \citep[][]{Stetson-1987}. The seeing of the images used in this work was on average of order $0.8 \pm 0.2\arcsec$, and images with seeing poorer than $1.7\arcsec$ were discarded from further analysis. The individual images were aligned using the {\sc daomaster} routine, which computes the astrometric transformation equation coefficients from the {\sc daophot/allstar} results. All available images were combined using the {\sc montage2} routine to make the master image of the area. More details can be found in \citet{contreras-2017}. 

In this paper, we focused on VVV tiles b302, b305-b309, b319, b320, b364, and b365, which are covered by our PSF photometry and which present maximum overlap with the OGLE survey, in the sense that a minimum of 50 OGLE Miras are already known in each tile (Table~\ref{table:coo}). This choice allows us to check and refine our search procedures, which will be extended to other VVV tiles in the near future. The OGLE data were taken with the 1.3\,m Warsaw telescope at Las Campanas Observatory, northern Chile, in the course of the OGLE-III\footnote{As of this writing, OGLE-IV data \citep{Udalski-2015} for bulge LPVs have not yet been made publicly available.} campaigns \citep[][]{Udalski-2003,Soszynski-2013}, and are available online.\footnote{\url{ftp://ftp.astrouw.edu.pl/ogle/ogle3/OIII-CVS/blg/lpv/}} Our resulting LPV catalog covering these ten tiles is presented in this paper; analysis of the remaining VVV tiles is currently in progress and will be the subject of forthcoming papers. 

Figure~\ref{Fig:map} provides a schematic map of the innermost region of the MW, with the locations of the centers of VVV tiles indicated by their respective IDs and the region for which our PSF photometry is completed indicated by the large rectangle and grid at $-10.0 <\ell({\rm deg})< +10.0$, $-2.5 < b({\rm deg}) < +2.8$. In this plot, red dots represent stars classified as Miras by the OGLE-III project. As can be readily inferred from this plot, tiles b302, b305-b309, b319, b320, b364, and b365 (highlighted in blue) are those with the largest number of previously known Miras, which is the reason why we specifically select them for this first paper in this series (see Table~\ref{table:coo}). According to our PSF photometry, these tiles comprise an initial sample of 61,627,100 individual sources with at least 20 VVV $K_s$-band observations each, which is the minimum number we adopt to ensure we only use sufficiently well-sampled light curves in our analysis.

\section{Variable star search}\label{sec:search}

In our approach, and in order to reduce the impact of statistical fluctuations that may impact some indices more than others, we use a combination of both correlated and noncorrelated statistical indices \citep[see][]{FerreiraLopes-2015wfcam,FerreiraLopes-2018papIII,FerreiraLopes-2016papI,FerreiraLopes-2017papII} to carry out our search for bona-fide variables. 
Specifically, our candidates were extracted according to the following procedure:

\begin{figure}
   \centering
   \includegraphics[width=0.45\textwidth,height=0.325\textheight]{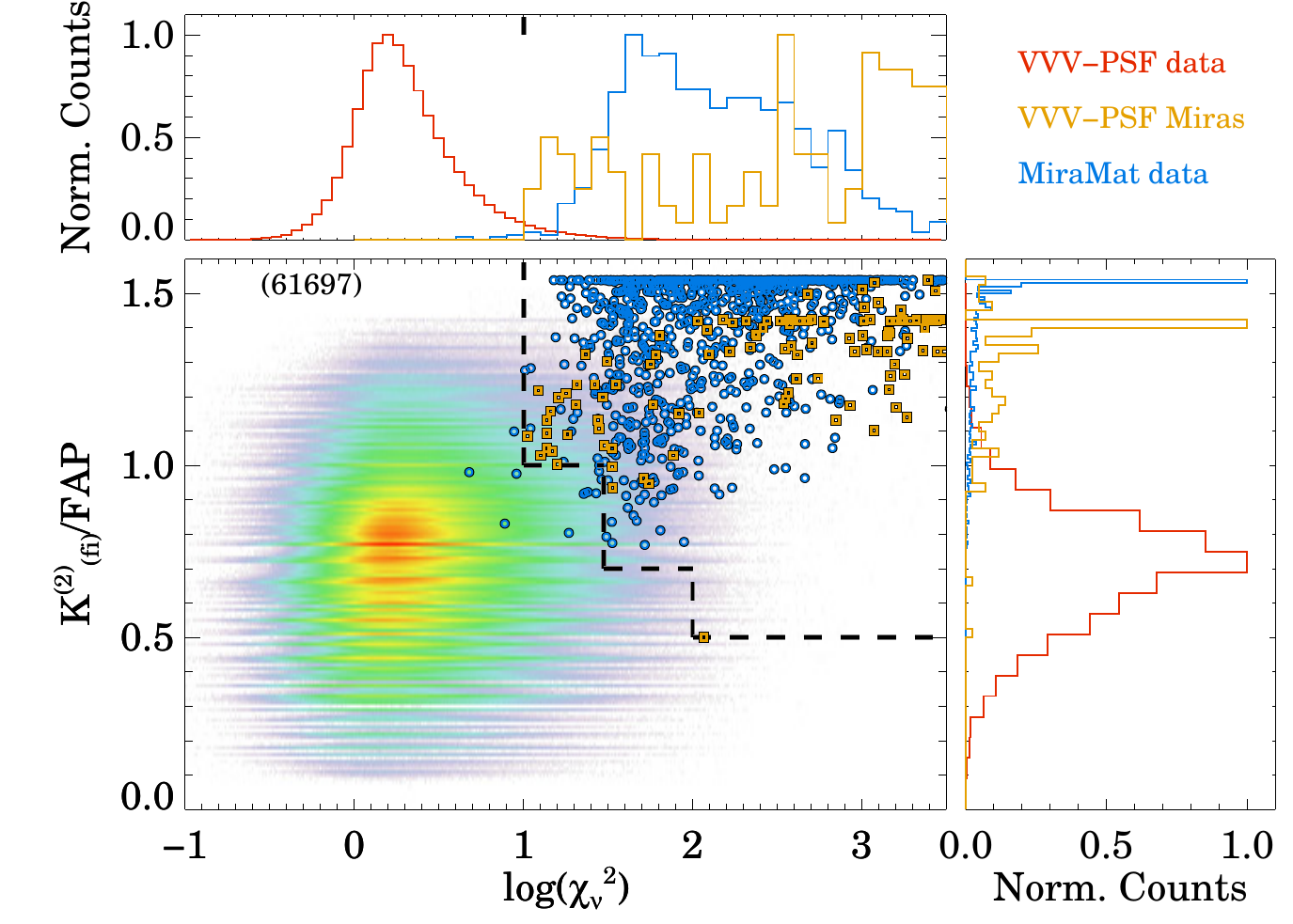}
   \caption{Distribution of VVV-PSF sources in the $K^{(2)}_{(fi)}$ vs. $\chi_\nu^2$ plane.
   The Miras identified by \citet[][labeled MiraMat in the plot]{Matsunaga-2009} are shown as blue circles. The region marked by dashed lines corresponds to the one from which we extracted our initial Mira candidates. The latter, upon further analysis and inspection, eventually led to our final selection of Miras, shown in this plot as orange squares.
   The maximum number of targets per pixel is 61,697 in this plot, corresponding to the dark orange-red region in the main panel.}
    \label{Fig:varindex}%
\end{figure} 

\begin{enumerate}
\item  Outlier removal: for each star, measurements outside five standard deviations from the median value were excluded from our analysis. In addition, measurements with unusually large associated error bars were also discarded. To accomplish the latter, we first computed the mean error bar and the corresponding standard deviation, using all available data points for the star. In those cases where the individual photometric error was higher than five standard deviations from the mean error bar, the measurement was discarded.
    
\item The reduced $\chi^2$ statistic, $\chi_\nu^2$, was computed with respect to a constant model \citep{Bellm-2021}. This quantity is defined as
    
   \begin{equation}
    \chi_\nu^2 = \frac{1}{N-1} \sum_{i=1}^{N}\frac{\left(m_i - \bar{m}\right)^2}{\sigma_{i}^2} ,
    \label{Eq:chi2}
   \end{equation}
   
   \noindent where $\bar{m}$ represents the weighted mean magnitude, $\sigma_i$ is the error associated with magnitude $m_i$, and $N$ is the number of valid observations.
   This is a commonly used noncorrelated index that is frequently employed to select Miras and other high-amplitude variables from photometric databases \citep[e.g.,][]{Carpenter-2001,Braga-2019}. 
\item  In parallel, the {\em correlated} index $K^{(2)}_{(fi)}$ was computed. It is defined as follows:  
    
    \begin{equation}
    K^{(2)}_{(fi)} = \frac{N^+}{N_C},
    \nonumber
   \end{equation}
    
    \noindent where $N^+$ is the number of positive correlations and $N_C$ is the total number of correlations, for an assumed $\Delta T$ value~-- the latter quantity corresponding to the time interval between observations. In our case, the correlation $\delta_{i,j}$ between a pair of measurements $i$, $j$ is defined as follows: 

\begin{equation}
   \delta_{i,j} = \left( \frac{m_{i}-\bar{m}}{\sigma_{i}} \right) \times \left( \frac{m_{j}-\bar{m}}{\sigma_{j}} \right) = \Lambda_{i,j} \,  \frac{|m_{i}-\bar{m}|}{\sigma_{i}} \times  \frac{|m_{j}-\bar{m}|}{\sigma_{j}},  
\label{delta_stet}     
\end{equation}

\noindent where $\bar{m}$ denotes the mean magnitude value, $\sigma$ the error bar, and $\Lambda_{i,j}$ reflects the sign of the multiplication, i.e., $+1$ or $-1$. According to this definition, $\delta_{i,j}$ describes the degree to which two measurements vary together in a time series. The $K^{(2)}_{(fi)}$ index is computed as the mean value over all ($N_C$) $i,j$ combinations, i.e., $K^{(2)}_{(fi)} = \sum \Lambda_{i,j}/N_C = N^+/N_C$. The $K^{(2)}_{(fi)}$ has a weak dependence on outliers, and an accurate estimation is obtained when only observations carried out close in time are considered, as done in this work \citep[for more details, see Sect. 3 in][]{FerreiraLopes-2016papI}.
While correlated indices in general, and the one used here in particular, are at least three times more efficient than noncorrelated indices \citep{FerreiraLopes-2015wfcam,FerreiraLopes-2018papIII,FerreiraLopes-2016papI,FerreiraLopes-2017papII}, they only work if certain conditions are satisfied \citep[][]{FerreiraLopes-2016papI}. In particular, the ratio between $\Delta T$ and the variability period $P$ should be smaller than $0.1$ \citep[$\Delta T/P << 0.1$; see Fig.~2 in][]{FerreiraLopes-2016papI}, in order to ensure accurate correlated values. To satisfy this criterion, in our work $\Delta T$ was increased until the number of correlated measurements reached $10$, and $\Delta T = 5$~d was set as an upper limit. This ensured $\Delta T/P << 0.1$, since Miras have periods typically in excess of $\sim80$ days. Indeed, $\Delta T$ can be quite large, if one considers the long periods typically found for LPVs. Such an approach allows us to compute the $K^{(2)}_{(fi)}$ index for about $97\%$ of the initial sample. \citet[][]{FerreiraLopes-2016papI} also defined an empirical relation to determine the expected value for the noise ($Fap_{K_{(fi)}}$, their Eq.~16), based on which unreliable index values can be identified. Indeed, those stars having periods shorter than $\Delta T$ can mimic noise according to those authors. Therefore, our approach improves the detection of LPVs and also reduces the number of stars having short periods. 

\begin{figure}
   \centering
   \includegraphics[width=0.45\textwidth,height=0.325\textheight]{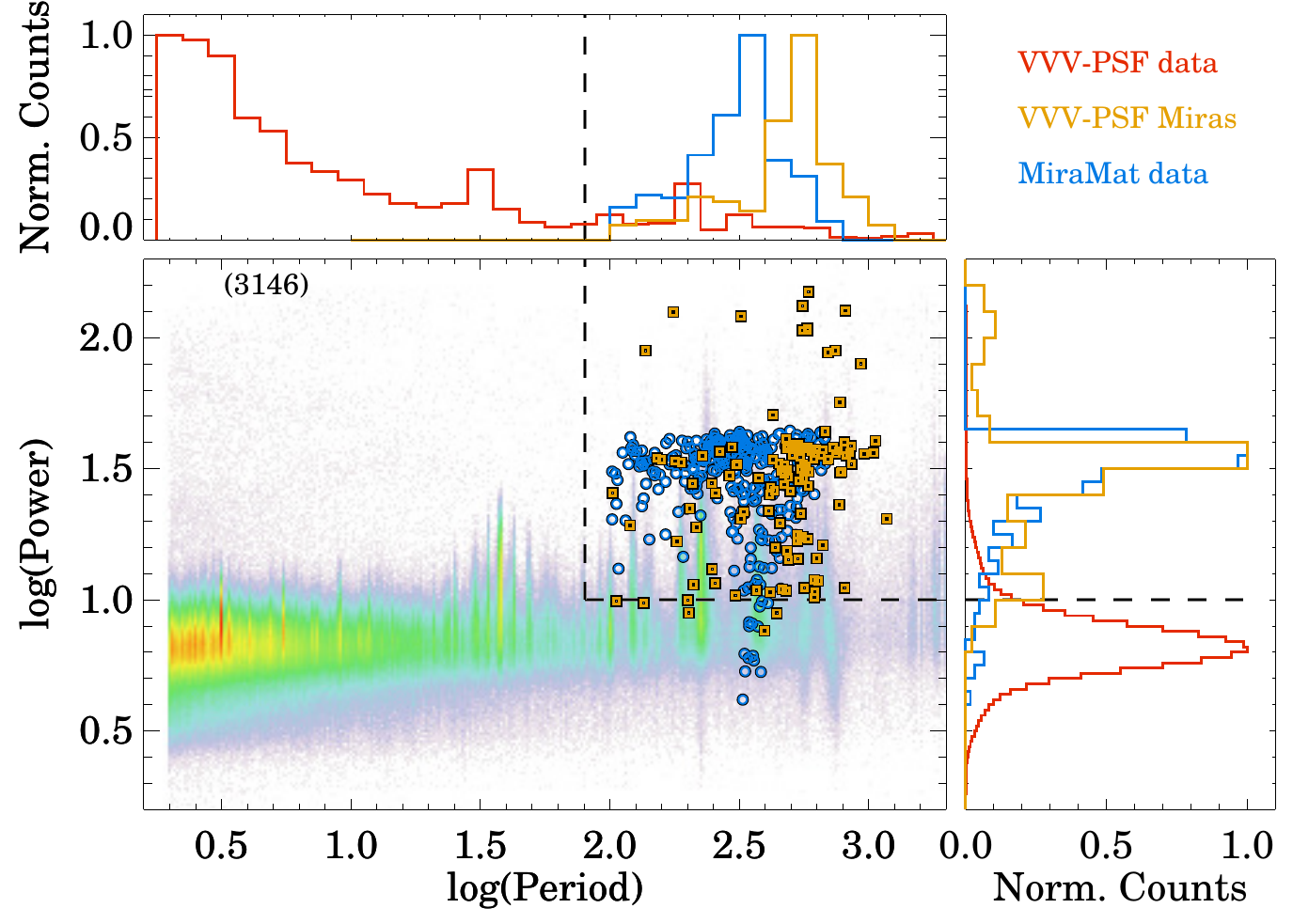}
   \caption{Logarithm of the power as a function of the period for the selected VVV-PSF sources (see region delimited by black dashed line in Fig.~\ref{Fig:varindex}). Symbols and colors are the same as in Fig.~\ref{Fig:varindex}) as well.
 The maximum number of targets per pixel is 3146 in this plot, corresponding to the dark orange-red region in the main panel.}            
   \label{Fig:periodpower}
\end{figure}

\item All sources having $K^{(2)}_{(fi)}/Fap$ and $\chi_\nu^2$ inside the region depicted by the dashed line in Figure~\ref{Fig:varindex} were then selected for further analysis. The borders of this selection region were adjusted manually in order to provide an optimum balance between the number of bona-fide LPV candidates and spurious sources (for more details, see Sect.~\ref{sec:catalog}). This leads to a preliminary sample having 365,696 candidates. The minimum cutoff in $\chi_\nu^2$ is smaller than typically used to select high-amplitude Miras, but was adopted in order to ensure detection of smaller-amplitude LPVs, as in the case of SRVs, and Miras whose amplitudes might be spuriously reduced in our photometry on account of saturation (see Sect.~\ref{sec:LPVsel}). On the other hand, the $K^{(2)}_{(fi)}/Fap > 1$ line is expected to enclose all nonsaturated LPVs.  
    
\item The periods of variable star candidates were estimated using three different methods, namely phase dispersion minimization \citep[PDM;][]{Stellingwerf-1978}, Lomb-Scargle  \citep{Lomb-1976,Scargle-1982}, and generalized Lomb-Scargle \citep{Zechmeister-2009}. A frequency interval between 0.5~cycles/day (corresponding to a period of 2 days) and $1/{\rm range}(t)$, where ${\rm range}(t)$ is the total time span in days, was adopted. In most cases, the periods found by the different methods were in good agreement. In the end, upon visual inspection of the results, the Lomb-Scargle solutions were finally adopted.
        
\item The periods thus estimated were used to compute the signal-to-noise ratio (S/N), defined as the ratio of the amplitude ($\Delta{K}$, measured peak-to-peak from the model) and the standard deviation of the residuals. The model was obtained using a Fourier series having three harmonics:
    
    \begin{equation}
    m(t) =  m_0+ \sum_{j = 1}^{3} a_{j}\cos\left( \frac{2\pi \, j \, t  }{ P }  + \theta_j \right) ,
    \label{eq_best_harm}
   \end{equation}

\noindent where $a_{j}$ and $\theta_j$ are the Fourier amplitude and phase coefficients, respectively, and $m_0$ is the star's mean magnitude \citep[e.g., ][]{Debosscher-2007,FerreiraLopes-2015cycles,FerreiraLopes-2015mgiant,Matsunaga-2017}.
    
\item The period and its respective power, together with the S/N estimated according to the procedure described in the previous step,
    were used to select our targets, which were required to satisfy one of the following criteria: \\
    
       \subitem (a) Power $> 20$ and S/N $> 1$;
       
       \subitem (b) $ 9 < $ Power $< 20$ and S/N $> 2$;
       
       \subitem (c) $ 5 < $ Power $< 9$ and S/N $> 3$;
       
       \subitem (d) Power $> 10$, S/N $> 1$, and $\langle K_{s} \rangle < 12$. \\

We note that criterion (d) includes only sources that did not pass criteria (a) or (b). In addition, a minimum period of $80$ days was also imposed. The power cutoff value is motivated by \citet{Horne-1986}, whose Eq.~13 indicates that, for a number of observations exceeding $\sim20$ epochs, a $Fap$ of $\sim20\%$, $\sim2\%$, $\sim0.2\%$, $\sim0.0\%$ results for the power used in the (a), (b), (c), and (d) criteria, respectively (see horizontal dashed line in Fig.~\ref{Fig:periodpower}). Compared with criterion (a), criteria (b) and (c) relax the constraint on the periodogram power, provided a sufficiently large S/N is still present in the light curve. Finally, compared to criterion (a), lower limits in the periodogram power and S/N were adopted for sources having many saturated observations ($\langle K_{s} \rangle < 12$) but still clearly defined variability, as saturated measurements can reduce the S/N in particular to values smaller than two. We emphasize that these criteria are sufficiently lax that bona-fide Miras, including fainter ones with large photometric errors, are unlikely to be missed at this stage, even if at the expense of a large number of contaminants. As a result, 12,476 sources were selected; we call these our VVV-PSF LPV candidates.
   
\item Given that, as already stated, the aforementioned criteria imply the presence of a significant number of contaminants, visual inspection was then performed, in order to remove the most obvious sources that are unlikely to be bona-fide LPVs. As a consequence, 11,333 of these 12,476 sources were considered to have variability signals that were either unclear or deemed unlikely to correspond to LPV stars. The remaining 1143 stars were finally assigned to two subgroups, namely ``Mira candidates'' (130 stars) and ``LPV$^{+}$ candidates'' (1013 sources). The latter correspond mainly to those cases where period determination was particularly problematic (for more details, see Sect.~\ref{sec:LPVsel}). Table~\ref{table:summaryselection} provides a summary with the number of selected targets according to each of the criteria mentioned above.

\begin{table}
\caption{LPVs selected according to the different criteria.} 
\label{table:summaryselection}      
\centering  
\begin{tabular}{l c c c }
\hline
\hline
  \multicolumn{1}{c}{Criterion} &
  \multicolumn{1}{c}{No. of Targets} &
  \multicolumn{1}{c}{Mira Cand.} &
  \multicolumn{1}{c}{LPV$^{+}$ Cand.}  \\
\hline
  \hline
     (a) & 4072 &  95 &   404\\
     (b) & 5550 &  28 &   455\\
     (c) & 1356 &  3  &   58\\
     (d) & 1498 &  4  &   96\\
\hline
Total & 12,476 & 130 & 1013 \\ 
\hline
\hline
\end{tabular}
\end{table}

\end{enumerate}

\subsection{Photometry recalibration}\label{sec:calib}

Our PSF photometry was originally calibrated to the VISTA photometric system by CASU \citep{gonzalez_fernandez2018}. CASU provides $ZYJHK_{s}$ aperture photometry catalogs directly calibrated from 2MASS point source data. We calibrated our instrumental PSF photometry in each chip through the comparison of common sets of stars with respect to CASU, which would place our photometry in the 2MASS photometric scale. However, as pointed out by \citet{Hajdu2020}, the standard CASU calibration procedure has some issues biasing the photometric scale, especially in regions of high stellar density. Therefore, we decided to recalibrate our PSF CASU-system photometry directly from 2MASS by carefully choosing the set of standards used in the procedure. We closely followed the general recommendations laid out in \citet{Hajdu2020}, but since we introduced a few modifications and additions, we briefly describe our procedure here.

For each chip catalog, we select stars with $JHK_{s}$ magnitudes (according to our PSF CASU-tied photometry) in the range of 12.0-15~mag. This range was adopted because we want to find matches in an equivalent (but shorter) range of 2MASS magnitudes, keeping in mind the fact that our zero points might still be wrong by up to $\pm 0.2$~mag or so at this stage \citep{Hajdu2020}. Indeed, the presence of these potential errors in the zero points are the motivation behind this recalibration. We remove all sources with companions closer than 3\arcsec (corresponding to the mode value of the 2MASS PSF), unless the companions are all fainter than 8 times the faint limit of the initial selection. This eliminates sources which might be blended with a significant (in terms of brightness) contaminant in 2MASS, and completes the definition of our initial catalog. Next, we cross-match this VVV selection with the 2MASS point source catalog, adopting a stringent radius of 0.4\arcsec. We keep all sources with 2MASS photometric quality flag ``AAA,'' with color $J-K_{s}<2$, and with an extinction-corrected color of $0.0<J-K_{s}<1.0$, according to the 2MASS photometry (see below for a discussion of the adopted extinction values).

A few more cuts are needed to define the calibration sample. They require to transform the retrieved 2MASS photometry to the VISTA photometric system. To this end, we adopt the version 1.5 set of transformation equations defined by CASU:

\begin{equation}\label{eq:2m-vvv-j}
J_{\rm 2M-VVV} = J_{\rm 2M} - 0.031 \, (J_{\rm 2M} - K_{\rm 2M})\,, \\
\end{equation} 
\begin{equation}\label{eq:2m-vvv-h}
H_{\rm 2M-VVV} = H_{\rm 2M} + 0.015 \, (J_{\rm 2M} - K_{\rm 2M})\,, \\
\end{equation} 
\begin{equation}\label{eq:2m-vvv-k}
K_{s, {\rm 2M-VVV}} = K_{\rm 2M} - 0.006 \, (J_{\rm 2M} - K_{s, {\rm 2M}}) \\
               + 0.005\, E(B-V)_{\rm corrected}, 
\end{equation} 

\noindent where 2M stands for 2MASS. 
As it can be seen, we need a color excess term, $E(B-V)_{\rm corrected}$, to compute the $K_{s}$-band transformation. Following \citet{gonzalez_fernandez2018}, we retrieve $E(B-V)$ at the coordinates of each star from the extinction map of \citet{schlegel1998}, and in all cases where the extinction-corrected $(J-K_{s})$ color of the star is $\leq0.25$, we recompute the extinction $E(B-V)_{\rm corrected}$ by imposing $(J-K)_0=0.5$. We note that, since the extinction term in Eq.~\ref{eq:2m-vvv-k} is very small, the impact of this choice upon the derived magnitudes is exceedingly small in $K_s$ and, according to Eqs.~\ref{eq:2m-vvv-j} and \ref{eq:2m-vvv-h}, strictly zero in the cases of $J$ and $H$ \citep[see also][]{Hajdu2020}.  

To avoid VISTA nonlinearity and/or saturation issues for bright targets, at this stage we remove stars with  $J,H,K_{s}<12.5$, according to their 2MASS magnitudes converted to the VVV system \citep[see also][]{ContrerasRamos-2017}. This completes the selection of the calibration sample. However, if the several applied cuts yield a sample of less than 25 stars, we remove the color cut on 2MASS photometry; if this is still not enough, we remove the cut on the reddening-corrected color as well.

Finally, the photometric zero-points in each of the $J$, $H$, and $K_s$ bands are computed from the median of the residuals of the respective $mag_{\rm 2M-VVV}-mag_{\rm VVV}$ distribution. The zero-points are added to the input original PSF magnitudes to cast them into the corrected 2MASS system. On the other hand, if the calibration sample had less than 3 stars, no zero-points were computed for that chip's catalog. We arrived at that situation only in a handful of cases, and therefore, for these chips, we interpolated zero-points from the respective $(\ell,b)$ zero-point maps, as the spatial variations of the zero-points are sufficiently smooth.

\subsection{LPV selection}\label{sec:LPVsel}

\begin{figure*}
   \centering
   \includegraphics[width=0.75\textwidth,height=0.4\textwidth]{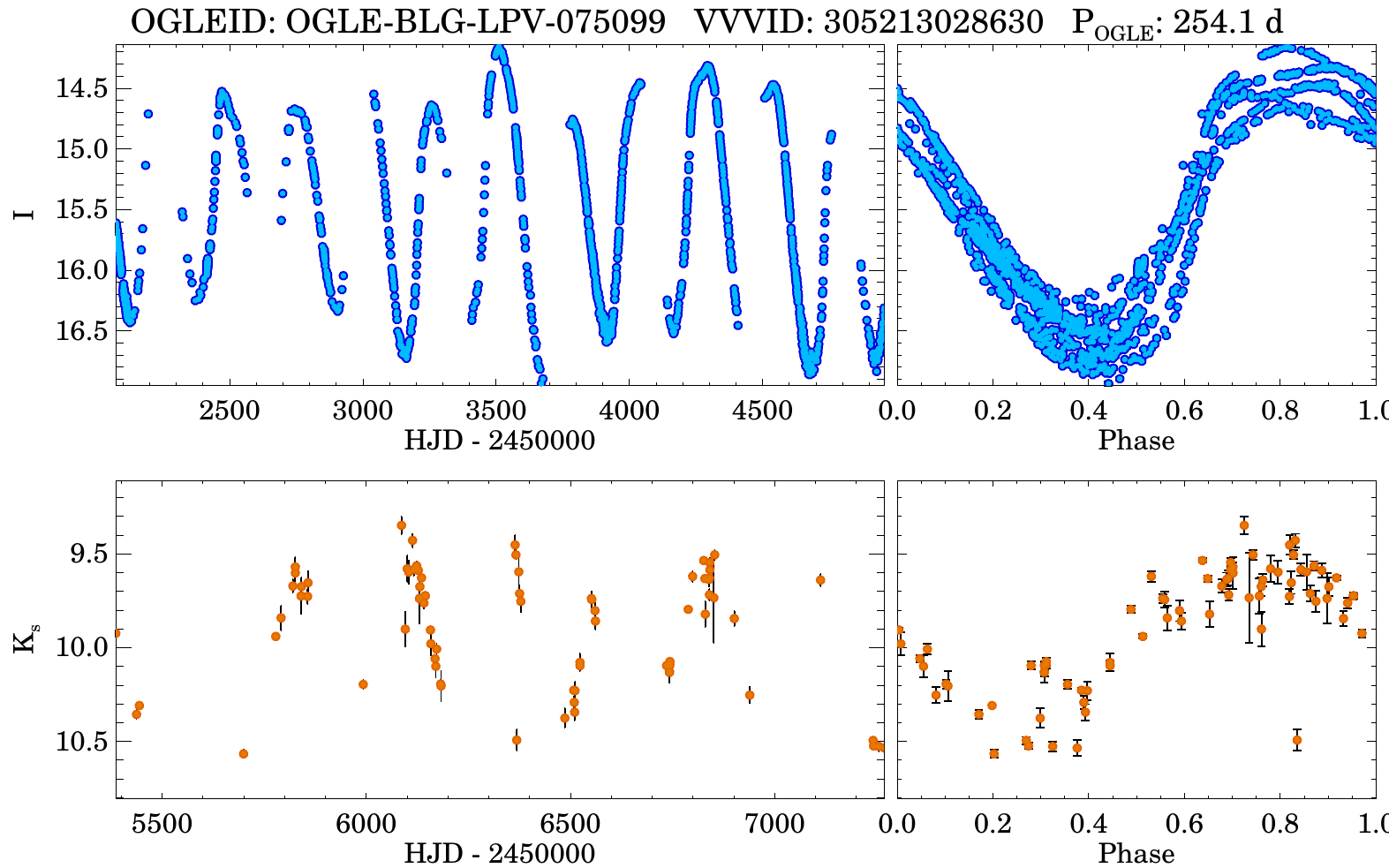}
   \caption{Example of raw (left) and phased (right) light curves, for a source cross-matched between OGLE (top panel) and VVV (bottom panel). The OGLE and VVV IDs, as well as the period found by OGLE, are displayed in the top panel.}
    \label{Fig:oglelcs}%
\end{figure*}

Steps 1 to 4 in the procedure described in the beginning of this section were applied in order to ensure the recovery of the majority of the LPVs contained in our initial sample. About $95\%$ of the OGLE and \citet[][]{Matsunaga-2009} Miras follow these criteria. As a result, $0.6\%$ of the entire VVV initial sample was selected as candidates (see Fig.~\ref{Fig:varindex}). Steps 5 to 8 were then applied to evaluate the reliability of the variability signal. We note that, since LPVs are intrinsically very bright stars, many fall in the nonlinear regime of the VIRCAM detectors, and thus turned out to be saturated in the VVV data. We opted {\em not} to reject all of these saturated sources \citep[i.e., with $K_s \lesssim 12$ mag;][]{ContrerasRamos-2017}.
While our PSF photometry (Sect.~\ref{sec:obsdata}) allowed us to recover and measure these saturated LPVs, the photometry for these sources (and hence their reported mean magnitudes and amplitudes) should of course be treated with extreme caution. On the other hand, we have found that their variability periods can often still be properly recovered (an example is shown in Fig.~\ref{Fig:oglelcs}), which is the main reason we include them in our catalog. Saturated sources with unclear periodicity signals, on the other hand, were removed from our sample. We note that the ages of even the saturated stars can still be estimated, even when reliably measured mean magnitudes are not available, as long as their periods can still be computed on the basis of the available measurements (see Sect.~\ref{sec:ages} below). We encourage follow-up studies of these sources, so that their photometric properties can also be properly established, and thus estimates of their distances obtained. We note that numerous LPVs brighter than $K_s = 10$ are known to exist in the direction of the Milky Way bulge; in particular, \citet{Qin-2018} show, using 2MASS data, that they cover the range $K_s \approx 4 - 10$, and peak around $K_s \approx 7.5$ \citep[see also][who arrived at similar conclusions using 2MASS and DENIS near-IR measurements]{Groenewegen-2005}. The vast majority of these very bright Miras are inaccessible to VVV, and thus remain outside of the scope of this series of papers.

An amplitude criterion was then applied in order to distinguish Miras. In the visual, Miras are defined as LPVs whose amplitudes exceed 2.5~mag in $V$. A similar threshold is however not well-established in the near-IR, although the majority of Miras have been shown to have $K$-band amplitudes exceeding 0.4~mag \citep[][]{Whitelock-2000}. \citet[][]{Matsunaga-2009}, in turn, used amplitude cutoffs of 0.4~mag in any of $J$, $H$, or $K_s$ (depending on data availability) in their working definition of Miras. In this work, we similarly used a cutoff of 0.4~mag in the $K_s$ band to distinguish between our Mira and SRV candidates. Finally, a visual inspection of the resulting light curves was carried out, which led to the rejection of a number of non-LPV contaminants. 

Our thus derived catalog of LPVs is comprised of two subgroups, namely Mira candidates (130 sources) and LPV$^{+}$ candidates (1013 sources). The latter group includes all  variable sources satisfying the aforementioned criteria and whose variability is thus likely to be real, but for which the available data did not allow us to obtain fully convincing periods, some of which are likely to be multiperiodic. The LPV nature of some of these stars~-- i.e., whether they are Miras, SRVs, SRVs with long secondary periods \citep[][and references therein]{Soszynski-2021}, or other types of objects~-- is thus not conclusively settled in this study. In fact, a number of SRVs may potentially also be present among the Mira candidates; conversely, a number of Miras may be present in the LPV$^{+}$ sample, in addition to the OGLE-III Miras that are present there because of their unreliable VVV-PSF photometry. In this paper, even though we report all of these sources, analysis of the derived period distribution, ages, and distances is restricted to the Mira candidates sample (130 stars) only. 

The light curves of all our Mira candidates, plus a representative sample of LPV$^{+}$ candidates, are shown in Appendix~\ref{sec:appendix}. Their photometric properties are summarized in Appendix~\ref{sec:catalog}.

\section{Cross-identification and tests}\label{sec:crossmatch}

We cross-matched our initial sample of VVV-PSF LPV candidates, which as we have seen is comprised of 12,476 sources (see Table~\ref{table:summaryselection}), with the published OGLE-III Mira catalog \citep[][]{Soszynski-2013}, resulting in 202 sources in common. We next performed a visual inspection of their respective $I$- and $K_s$-band light curves. As a result, we find that the vast majority of these sources are saturated in the VVV photometry and present poorly defined light curves. The only OGLE source in our sample of Mira candidates is OGLE-BLG-LPV-075099, whose light curve is displayed in Figure~\ref{Fig:oglelcs}.

The fact that the majority of OGLE sources are saturated in VVV is not an unexpected result, for two main reasons. First, the VVV survey was specifically designed to go deeper in regions of high interstellar extinction than afforded by previous surveys, both in the visible (such as OGLE) and near-IR (such as 2MASS). Thus, sources in the VVV footprint that are bright in the visible can easily saturate the VIRCAM detectors. On the other hand, faint sources detected in the near-IR using VVV data will typically not be present in visual catalogs, as confirmed by our analysis. 

Whenever available, OGLE periods were used, as they are based on much longer and robust datasets, where the periodicity is much more clearly defined than in our study (see Fig.~\ref{Fig:oglelcs}). In any case, for the one source in common, our period (255.8~d) and the one from OGLE (254.1~d) differ by only 1.7~d. Figure~\ref{Fig:oglelcs} shows one cross-matched source of our initial sample with the published OGLE-III Mira catalog. While this source is saturated in the VVV $K_s$-band data, the cyclical nature of its variability can still be observed.

In addition, our sample was cross-matched with the Set of Identifications, Measurements and Bibliography for Astronomical Data 
\citep[SIMBAD;][]{Wenger-2000} and the American Association of Variable Star Observers (AAVSO) Variable Star Index \citep[VSX;][]{Watson-2014}. In total, $43$ matches were found, with those sources being classified as Mira candidates ($2$ sources), asymptotic giant branch stars (AGB,  $13$ sources), and infrared sources ($28$ sources). We also cross-matched with the {\em Gaia} DR2 LPV catalog \citep[][]{Mowlavi-2018}.
Positive matches mainly correspond to saturated sources in our VVV-PSF photometry. However, there are two {\em Gaia} DR2 matches that are {\em not} saturated in our photometry, and whose periods also fall in the right range for the computation of photometric distances; these will be further discussed in Sect.~\ref{sec:distance} below. Lastly, we cross-matched our sources with known clusters located in our studied fields, which includes four globular clusters (Terzan~9, Djorg~2, Terzan~10, and NGC~6544) and a number of open clusters \citep{Dias-2002}. No cluster members could be confirmed in our analysis.

\begin{figure*}
   \centering
   \includegraphics[width=0.75\textwidth]{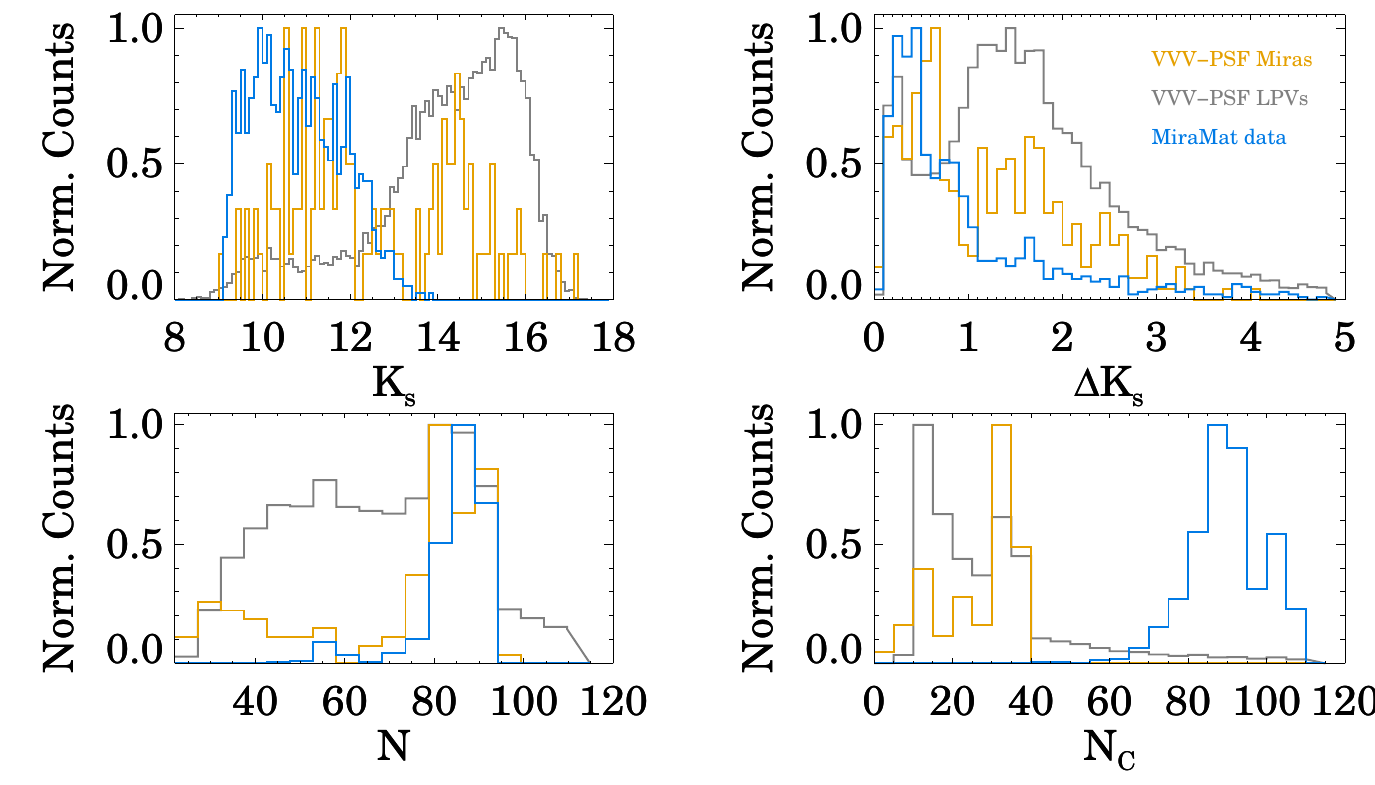}
   \caption{Comparison of VVV-PSF candidates with the sample of Miras from \citet[][]{Matsunaga-2009}. Histograms are shown of mean $K_s$-band magnitudes (upper left), amplitudes $\Delta{K_{s}}$ (upper right), number of observations $N$ (bottom left), and number of correlated measurements $N_C$ (bottom right). The VVV-PSF Miras, MiraMat sample, and VVV-PSF LPV candidates are shown in blue, yellow, and gray, respectively.}
    \label{Fig:histmiramat}%
\end{figure*}

\begin{figure*}
   \centering 
   \includegraphics[width=0.455\textwidth]{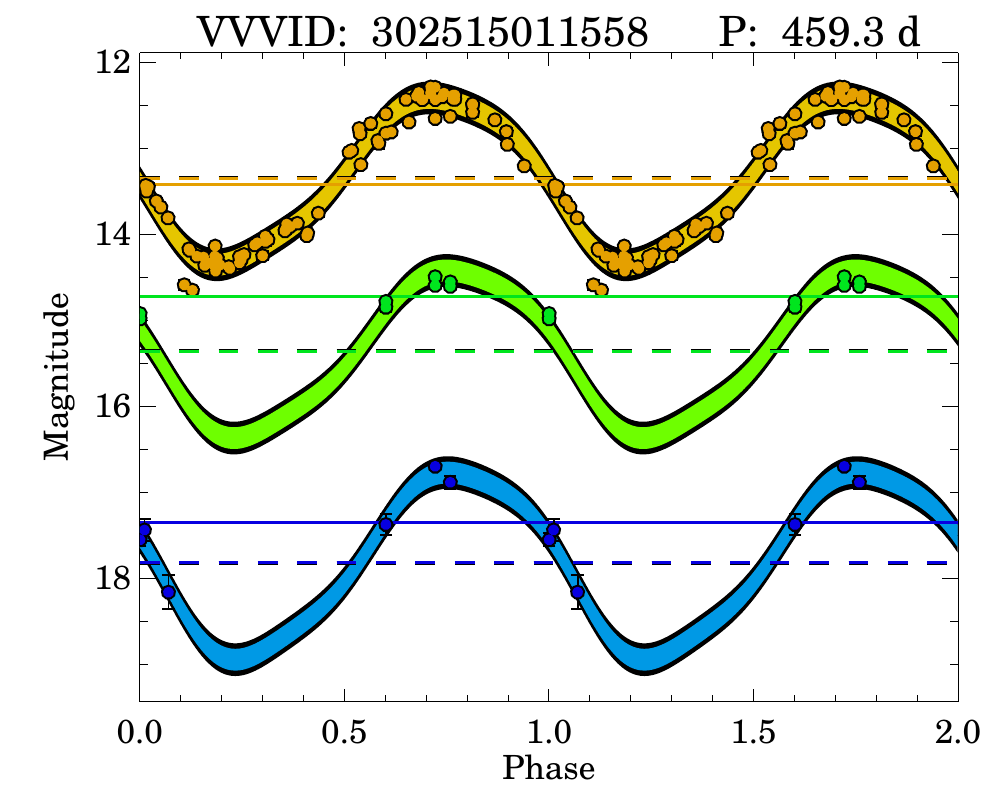}
   \includegraphics[width=0.45\textwidth]{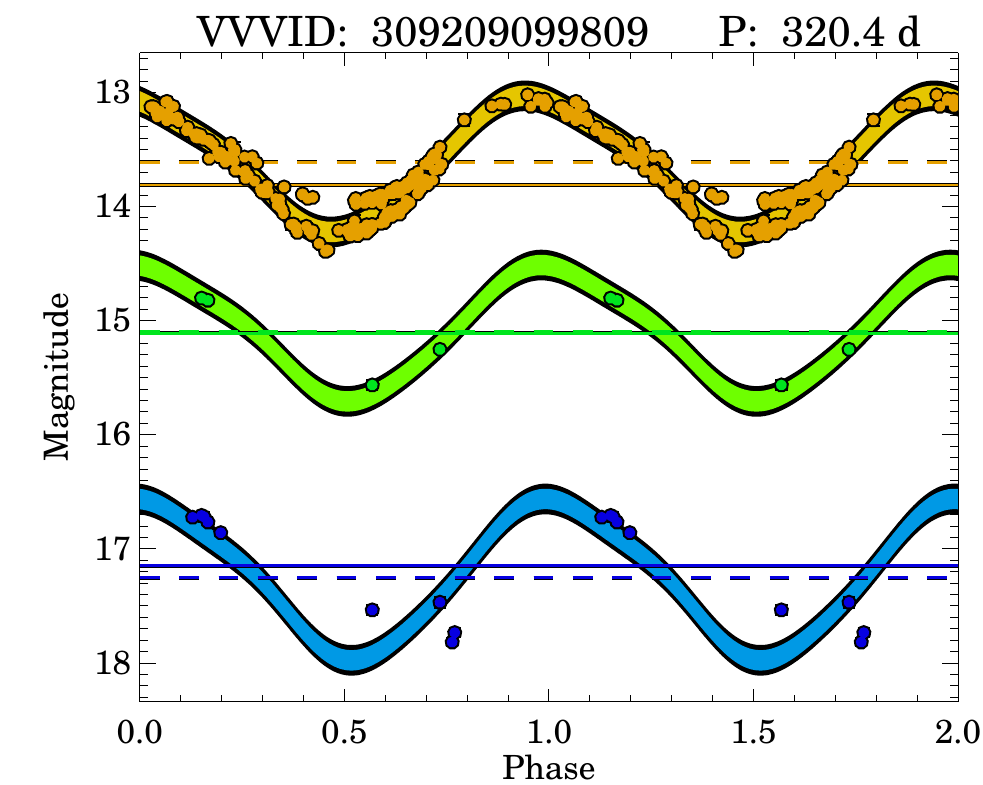}
   \includegraphics[width=0.45\textwidth]{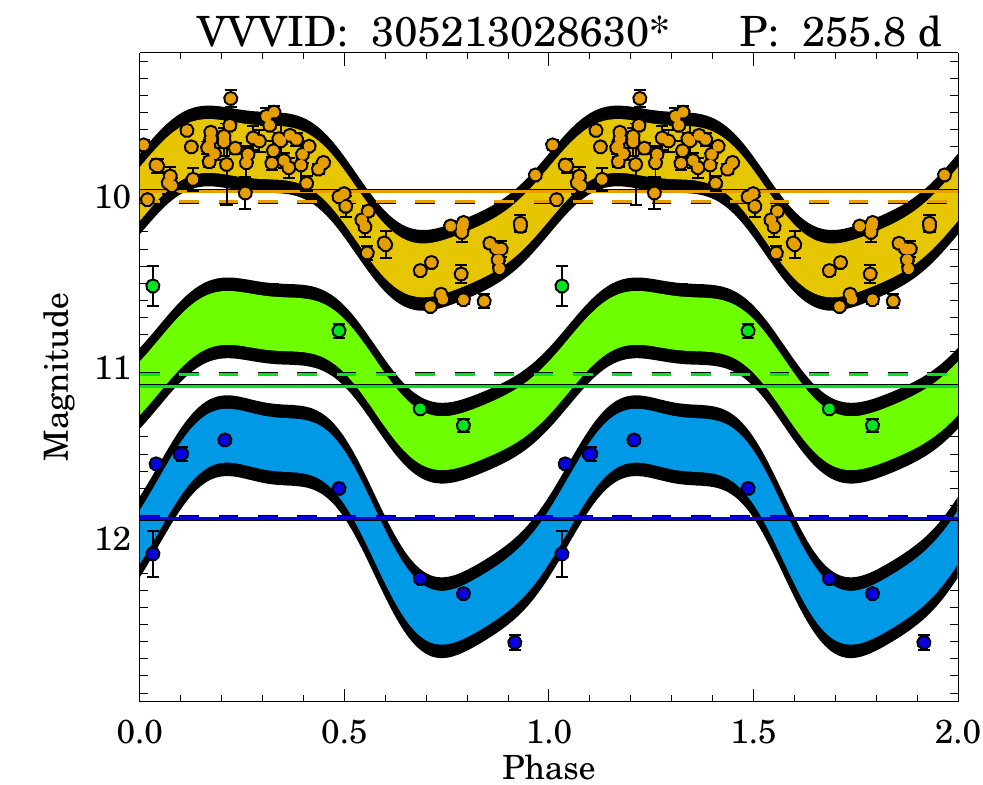}
   \includegraphics[width=0.45\textwidth]{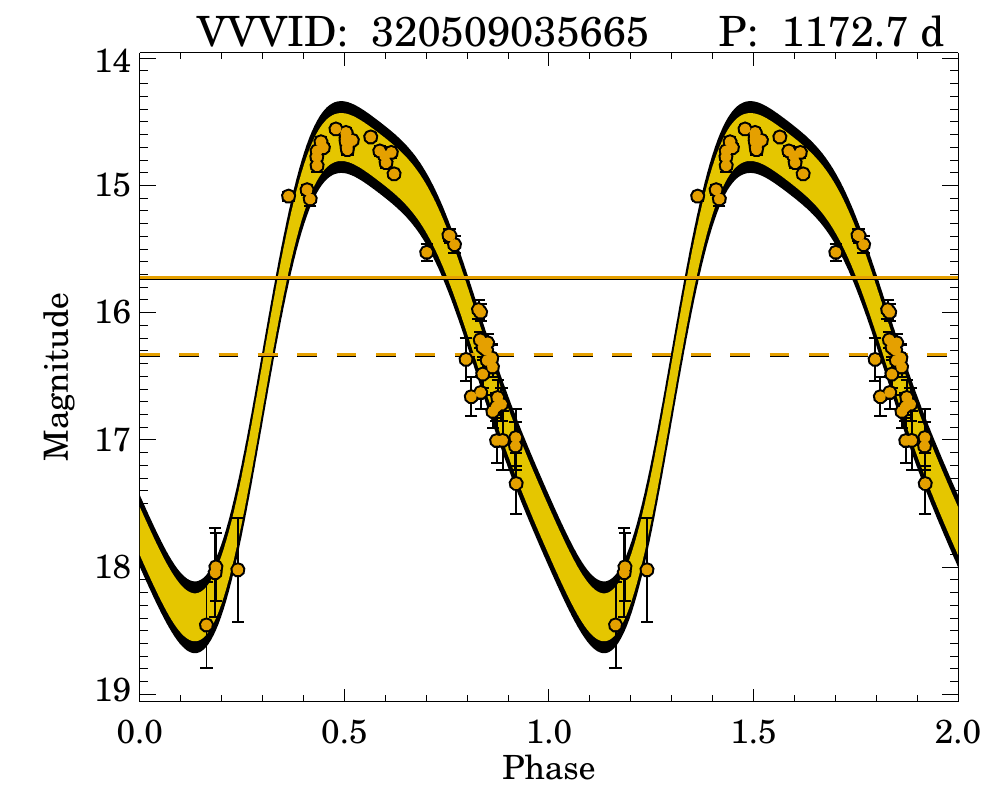}
   \caption{Examples of folded light curves in $J$ (blue), $H$ (green), and $K_s$ (yellow). The solid and dashed lines represent the mean magnitudes considering a straight average over the available observations and best-fit template model (if available), respectively. The color bands around data points correspond to a $2\sigma$ range around the mean. The VVV-ID and periods are included on top of each panel. In the bottom right panel, no data are available in the $J$ and $H$ bands, and so only the $K_s$ photometry shown. We note that distances could be computed for both stars in the upper row, but not for those in the bottom row, due to saturation ({\em bottom left panel}) or a period that places the star outside the distance indicator regime ({\em bottom right panel}).}
   \label{Fig:features}
\end{figure*}

The significant number of outliers in the partially saturated VVV-PSF light curves (see Figure~\ref{Fig:oglelcs}, bottom panels) for the OGLE cross-matched sources and the different passbands employed hinder the analysis of the criteria we used to select Miras. Therefore, the catalog of Miras with $K_s$-band light curves obtained by \citet[][hereafter MiraMat]{Matsunaga-2009} was also added to check our approach. This catalog contains $936$ Miras with reliable photometry. Next, variability indices, periods and feature parameters were computed using the same algorithms used to analyze the VVV-PSF data (Sect.~\ref{sec:search}). We find that, among the MiraMat sources flagged in their study as having good light curves and for which the authors could provide periods, about 98\% are selected, and their estimated periods are in agreement with those of \citet[][]{Matsunaga-2009}. 

Histograms of selected properties of MiraMat and VVV-PSF Miras are presented in Figure~\ref{Fig:histmiramat}. Additional comparisons are provided in Figures~\ref{Fig:varindex} and \ref{Fig:periodpower}. Based on these, the following main conclusions can be reached: 
 
\begin{itemize}              
     \item The VVV-PSF Mira magnitudes are distributed in the range of $8-16$~mag in $K_{s}$. The MiraMat sample does not reach as deep, covering $9-14$~mag in $K_{s}$.
     
     \item The number of correlated measurements ($N_C$) found for the MiraMat sample is at least twice that found in the VVV-PSF Miras, which shows the different observation strategies of these two surveys, keeping in mind that the total numbers of observations ($N$) are not markedly different.

     \item According to Figure~\ref{Fig:varindex}, about $50\%$ of the MiraMat sample stars assume values along the upper limit of the $K^{(2)}_{(fi)}/Fap$ distribution. 
    As expected, higher $N_C$ values lead to a better discrimination of Miras from noise.
    
     \item We caution that the analysis of the $\chi_\nu^2$ values for the MiraMat sample is biased, since the magnitude errors are unavailable (Fig.~\ref{Fig:varindex}). The majority of MiraMat sources are bright stars that are not saturated in their data, which can imply error bars smaller than adopted in our analysis ($\sigma_{i,{\rm MiraMat}} = 0.01$), and hence larger $\chi_\nu^2$ values according to Eq.~\ref{Eq:chi2}. On the other hand, the faint sources could have lower $\chi_\nu^2$ values if their corresponding observational error bars are larger than assumed in our calculations.
     
     \item Our VVV-PSF Mira candidates cover a wider range in $\chi_\nu^2$ than do the Miras in the MiraMat sample. In particular, the former sample has proportionally more stars with lower $\chi_\nu^2$ values than the latter.
 \end{itemize}
 
In summary, the fact that $\sim 98\%$ of MiraMat sources are successfully recovered and that our periods are in agreement with those from \citet[][]{Matsunaga-2009} lends support to our using our proposed methodology (Sect.~\ref{sec:search}). On the other hand, we caution that, while application of our method to the MiraMat sample demonstrates that it can be very successful with high-quality photometric data such as provided in their study, the method may not necessarily perform equally well when applied to different datasets, having different cadences, time coverage, total number of measurements, and photometric error distributions. To account for such differences, and depending on the desired purity of the sample to be selected prior to the visual inspection stage, the selection criteria laid out in Sect.~\ref{sec:search} may need to be tweaked accordingly.

\section{Light curve features}\label{sec:features}

Mean $K_s$ magnitudes and amplitudes ($\Delta{K_s}$) were estimated based on the models given by Eq.~\ref{eq_best_harm}. Indeed, in the $K_{s}$ case, the relative difference between a straight average over the available magnitudes (circles in Fig.~\ref{Fig:features}) and mean magnitudes computed on the basis of detailed light curve fits (colored bands in Fig.~\ref{Fig:features}) is on average $\sim1.8\%$. The largest differences are found when the available observations do not cover uniformly the whole light curve~-- for instance, when most data are placed close to maximum or minimum light. In the case of the star shown on the bottom right panel of Figure~\ref{Fig:features}, for instance, the relative lack of measurements close to minimum light biases the mean $K_{s}$ value, which is significantly brighter than found from the detailed light curve fit.

VVV-PSF measurements in $J$ and $H$ are not available for all sources in our sample (Table~\ref{table:master}). In those cases, we retrieved 2MASS magnitudes and examined both 2MASS and VVV images, in the hope of replacing the missing VVV values with 2MASS ones. Unfortunately, however, we concluded that for no such star was a reliable 2MASS magnitude available, especially due to the faintness of the stars and/or the presence of nearby companions,  unresolved in the 2MASS images.  

The VVV-PSF sample includes typically around three observations in the $J$ and $H$ bands, as compared to over $70$ in the $K_s$ band. Therefore, we can precisely determine mean magnitudes $\langle K_s \rangle$ and amplitudes $\Delta{K_s}$ using the available data, but a more careful approach is needed in the case of $J$ and $H$, since the available observations are generally insufficient to completely define the light curves. In order to obtain reliable $J$ and $H$ mean magnitudes and amplitudes, we resort to describe the $J$ and $H$ light curves by means of suitably scaled templates obtained initially using the $K_{s}$-band data. Our approach is based on the empirical relations for amplitudes and phase lags given by \citet[][]{Yuan-2018}, and the results are illustrated in Figure~\ref{Fig:features}. 

More in detail, we initially obtain a $K_s$-band template for each star, fitting the coefficients of the model given by Eq.~\ref{eq_best_harm} to our $K_s$ data. From the thus derived $K_s$ templates, we then obtain the corresponding templates in $J$ and $H$ by using the empirical relations for amplitude ratios and phase lags provided by \citet[][]{Yuan-2018}. The latter authors assume a linear dependence of these two quantities on the log-period; the corresponding coefficients are provided in their Table~2. Errors are unfortunately not available for these coefficients. However, their Figure~2 shows that the scatter around the values inferred from such linear expressions is of order $\approx \pm0.5$ in the amplitude ratios and $\approx 0.05$ in the phase lags ($1-\sigma$ level). By allowing the amplitude ratios and phase lags to vary within these ranges, we finally adopted, for each individual star and bandpass, the combination of amplitude ratio and phase lag that minimized the residuals of the fits to our data. The thus-fitted templates allow us to obtain reliable mean $J$ and $H$ magnitudes, along with the corresponding amplitudes, for all our stars with at least one $J$ or $H$ VVV-PSF measurement. Examples of our resulting template fits are shown in Figure~\ref{Fig:features}, which also indicates the resulting mean values. 

We note that, in our procedure, the coefficients corresponding to both C-rich and O-rich Mira provided by \citet[][]{Yuan-2018} were initially used. However, since the Mira population found in the Galactic bulge is mainly comprised of O-rich Miras \citep[][]{Soszynski-2013-2}, we finally decided to adopt the solutions corresponding to the latter. However, as we shall see later, a significant fraction of the Miras for which we could compute distances are not located in the bulge proper, and thus the assumption of an O-rich composition is not necessarily valid for those in particular. In this sense, Guo et al. (in prep) used VVV-PSF photometry from VIRAC2 \citep[][and in preparation]{Smith-2018} to perform a search for young stellar objects (YSOs) in the VVV footprint, independently finding a large number of Mira candidates. In those tiles in common with our analysis, they were able to recover most of our candidate Miras; in addition, based on WISE \citep[Wide-field Infrared Survey Explorer;][]{Wright-2010} mid-IR photometry, and using diagnostic color-color plots as in \citet{Lian-2014}, they were able to confirm that the majority of these Miras are indeed O-rich, even though there are a few that may be C-rich. Follow-up spectroscopy would be of significant interest to verify their color-based classifications. C-rich stars in the bulge are of particular interest as they may either belong to a young population or be the result of mergers \citep[][]{Feast-2013, Matsunaga-2017}. A full discussion of this topic is beyond the scope of this paper, and will be presented in Guo et al. (in prep).

\section{Results}\label{sec:catalog}

Appendix~\ref{sec:catalog} reports the main measured properties for the stars in our final catalog, containing, as we have seen, 130 Mira candidates and 1013 LPV$^{+}$ candidates. 
Columns 1 through 9 in Table~\ref{table:master} list, respectively, our Mira sources' IDs, coordinates, mean magnitudes, and amplitudes in each of $J$, $H$, and $K_s$, in addition to the period (in days), which is given in column 10. In columns 1 through 8 of Table~\ref{table:master2}, in turn, one can find a list of our LPV$^{+}$ sources' IDs, OGLE~IDs, coordinates, mean magnitudes in each of $J$, $H$, and $K_s$, and amplitudes in $K_s$, respectively.

\subsection{Period distribution}\label{sec:periods}

\begin{figure}
   \centering
   \includegraphics[width=0.95\hsize]{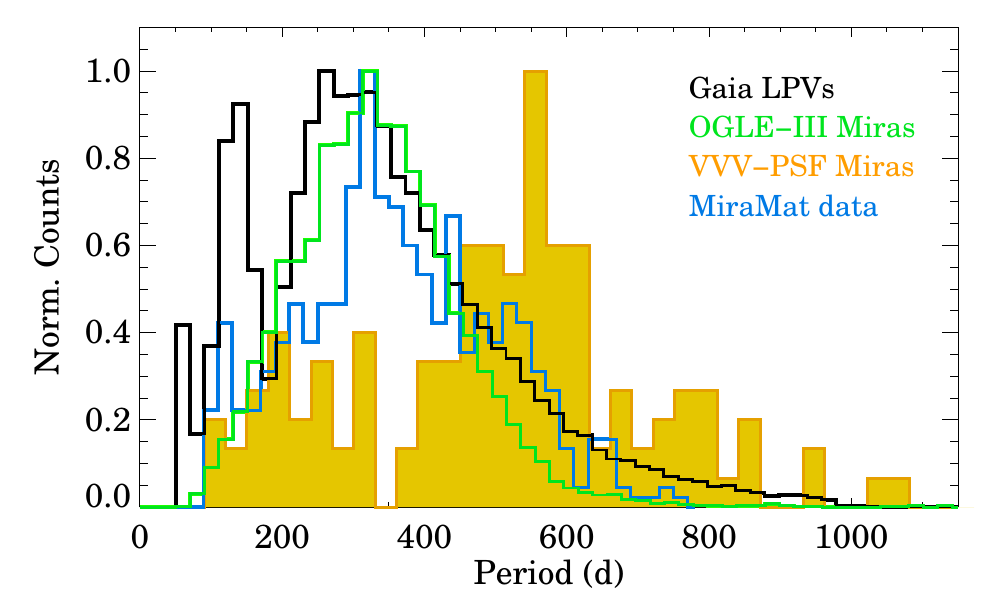}
   \caption{The period histograms of {\em Gaia} LPVs (black), OGLE-III Miras (green),  VVV-PSF Miras (orange), and MiraMat Miras (blue) are presented.}
   \label{Fig:his-p}%
\end{figure}

Our main goal in this work is to detect Mira stars. The period histograms of our VVV-PSF Miras, along with the Miras of the OGLE-III and MiraMat surveys, as well as {\em Gaia} LPVs, are shown in Figure~\ref{Fig:his-p}. Because of selection biases, the period histograms of these surveys are somewhat different. The derived period distributions depend on the observed Galactic regions (affecting the relative proportion of young versus old Miras), wavelength regime (bright and/or close versus faint and/or distant ones), and time span of the observation (short- versus long-period variables). For instance, the OGLE survey, as an optical experiment, is unable to detect Miras located in high-extinction areas, while being successful to cover Miras either situated quite close to us, or intrinsically young and bright. On the other hand, the {\em Gaia} survey, being an all-sky experiment, covers a comparatively broader sample of Miras, in terms of period range. We caution, however, that the {\em Gaia} LPV sample includes non-Miras as well, and so a more refined classification of the {\em Gaia} LPVs will be needed before stronger conclusions can be drawn.

In the case of the VVV survey, the brightest Miras are mainly saturated, therefore we do not identify their contribution in our sample unless they are located sufficiently far.
Even though we need to cover more tiles of the VVV survey to allow more detailed comparisons with other available surveys that cover wider fields, the bimodality of the VVV-PSF Miras histogram is readily apparent. More specifically, the VVV histogram reveals two notable peaks, around $\sim200$~d and $\sim 500$~d.
Other studies, in addition to those that are shown in the figure, have variously found a single peak similar to our long-period one \citep[][]{Whitelock-1991, Matsunaga-2005} or a bimodality that resembles ours to some extent \citep[][]{Glass-2001, Matsunaga-2009}, though with a short-period peak located at a longer period value than we find \citep[see also][]{Matsunaga-2009}. On the other hand, also using VVV data but with much smaller samples of Miras and SRVs than in the present study, \citet{Gran-2013} and \citet{Molina-2019} also found signs of bimodality in their period distributions, with minima located in the range $300 \lesssim P({\rm d}) \lesssim 400$.   

The bimodality that is seen in our sample thus indicates the presence of both young (long-period) and old (short-period) LPVs. \citet{Molina-2019} suggested that the demise of VVV LPVs in the $200-400$~d period range could be due to the fact that Miras located in the bulge proper would be saturated for periods of 200~d or longer, as they would have $\langle K_s \rangle \approx 11$ or less. Indeed, we see a minimum in the magnitude distribution of our Miras and Mira candidates at $\langle K_s \rangle \approx 11$ (see Fig.~\ref{Fig:histmiramat}), and have also verified that the bulk of the OGLE Miras that have cross-matches with our sources have $\langle K_s \rangle \approx 11$ {\em and} very poor VVV-PSF light curves. In addition, a selection bias may plausibly also be present, caused by the seasonal nature of VVV observations: we note that the minimum in our histogram is found close to 365~d (Fig.~\ref{Fig:his-p}), where most other non-VVV surveys have found significant numbers of LPVs. Whatever the exact reason(s) for this dip may be, more observations will be required in order to properly detect and establish the multiband light curve properties of intermediate-period LPVs located toward the Galactic bulge.  

\subsection{Period-amplitude relation}\label{sec:period-ampl}

\begin{figure}
   \centering
   \includegraphics[width=0.95\hsize]{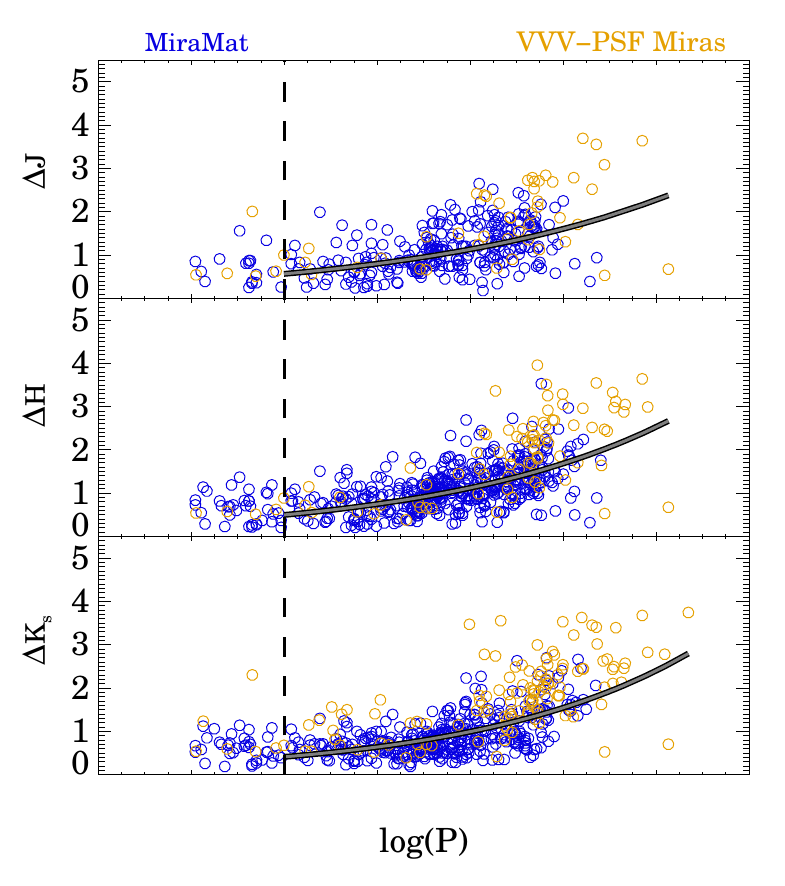}
              \caption{$\Delta J$, $\Delta H$, and $\Delta{K_{s}}$ amplitudes as function of logarithmic period, from top to bottom, respectively. The VVV-PSF Miras (orange circles) and MiraMat Miras (blue circles) are displayed. The dashed line marks the short-period limit of the \citet{Conroy-2015} relation (Eq.~\ref{eq_pamp}). The lines show our fits for the combined sample of VVV-PSF and MiraMat Miras. }
              \label{Fig:Avsp}%
\end{figure}

Figure~\ref{Fig:Avsp} indicates the $JHK_{s}$ amplitudes of VVV-PSF Miras as a function of their periods. The same parameters found in the \citet[][]{Matsunaga-2009} catalog are plotted as reference. The $\Delta J$, $\Delta H$, and $\Delta{K_{s}}$ amplitudes are in agreement with Matsunaga's results \citep[see also][]{Gran-2013}. As can be seen, the period-amplitude distribution of Miras gets tighter in the $K_s$-band, since the amplitudes are reduced in this bandpass. 

\citet[][]{Conroy-2015} introduced a period-amplitude relation in the $I$-band for Miras $(\log P>2.2)$:

  \begin{equation}
      \Delta I = 10^{\left(\alpha_I \log P + \beta_I\right)}  \,,
      \label{eq_pamp}
  \end{equation}
 
\noindent where $P$ is the period in days, $\Delta I$ represents the amplitude in the $I$ band, and $\alpha_I$ and $\beta_I$ are constants (see Table~\ref{table:amplitudecoefficients}). In the same fashion, we also establish $\alpha$ and $\beta$ for each of the $JHK_s$ wavebands. Our results are shown in Table~\ref{table:amplitudecoefficients}, using the combined sample of VVV-PSF and MiraMat Miras. 

Our amplitude-period relation in the $K_s$ band can be compared with the similar ones that were introduced recently by \citet{Molina-2019}, which however are valid only for $\log (P/{\rm d}) > 2.6$. As a result, we find that our expression provides $\Delta{K_{s}}$ values that are in better agreement with their Eq.~(2) in the period range $ 2.6 < \log (P/{\rm d}) < 3.18$, which is also the range of applicability of our expression (see Fig.~\ref{Fig:Avsp}). This is not unexpected, as Eq.~(2) in \citet{Molina-2019} was also obtained by combining VVV and MiraMat amplitudes, and predicts amplitudes that are smaller than the ones based on VVV-only amplitudes in the long-period regime (which is also apparent from our Fig.~\ref{Fig:Avsp}. For further discussions on the implications of these results, we refer the reader to \citet{Molina-2019} and \citet[][]{Ita-2021}, and the references cited therein.     

\begin{table}
\caption{Coefficients of the $IJHK_{s}$ Period-Amplitude Relations\tablefootmark{(a)}} 
\label{table:amplitudecoefficients}      
\centering  
\begin{tabular}{l c c }
\hline
\hline
  \multicolumn{1}{c}{Bandpass} &
  \multicolumn{1}{c}{$\alpha$} &
  \multicolumn{1}{c}{$\beta$}  \\
\hline
\hline
  $I$   &      0.50       & $-1.25$         \\
  \hline
  $J$   & $0.76\pm0.07$ & $-1.91\pm0.19$ \\
  $H$   & $0.88\pm0.05$ & $-2.24\pm0.14$ \\
  $K_s$ & $0.96\pm0.05$ & $-2.49\pm0.14$ \\
\hline
\hline
\end{tabular}
\tablefoot{
\tablefoottext{a}{
See Eq.~\ref{eq_pamp}. The coefficients for the $I$ band were introduced by \citet[][]{Conroy-2015}, while $JHK_{s}$-band coefficients are obtained using VVV-PSF and MiraMat Miras (see Fig.~\ref{Fig:Avsp}).} 
}
\end{table}

\subsection{Distance indicators and extinction} \label{sec:distance}

Thirty-one VVV-PSF Miras fall in the period range $100 \lesssim P({\rm d}) \lesssim 350$, over which the near-IR period-luminosity relation of Miras is known to be linear, and they are consequently expected to be reliable distance indicators \citep[e.g.,][]{Feast-2002,Matsunaga-2009}. Among these, $13$ have $\langle K_s \rangle < 12$~mag, and are thus in the regime of saturation according to \citet[][see also Sect.~\ref{sec:calib}]{ContrerasRamos-2017}. We exclude these stars from further analysis, and we are left with a final sample of $18$ sources.

In what follows, we refer to stars satisfying these criteria as {\em distance-indicator Miras}. Follow-up photometry of the remaining $13$ Miras in this crucial period range is strongly encouraged, as reliable distances could be obtained for them, thus providing useful insight into the structure of the inner Milky Way, and indeed beyond.

The absolute magnitudes of the distance-indicator Miras in our catalog were obtained by means of the period-luminosity relation derived by \citet{Matsunaga-2009} in the Infrared Survey Facility (IRSF)/SIRIUS photometric system, suitably transformed into the VISTA system. This was accomplished by first converting the IRSF relations to their 2MASS counterparts following \citet[][]{Carpenter-2001}, and the latter to the VISTA system according to the prescriptions given in  \citet{gonzalez_fernandez2018}.
The thus transformed equations are as follows: 

 \begin{equation}
      M_{J} = - 5.82 - 2.88\,(\log P - 2.3)
      \,, \label{eq:plj}
 \end{equation}
  \begin{equation}
       M_{H} = - 6.53 - 3.18\,(\log P - 2.3)
      \,, \label{eq:plh}
  \end{equation}
  \begin{equation}
      M_{K_{s}} = - 6.89 - 3.55\,(\log P - 2.3) 
      \,, \label{eq:plks}
  \end{equation}

\noindent where $M_{J,H,K_{s}}$ is the absolute magnitude and the period $P$ is in days. Naturally, stars in the inner regions of the Milky Way can be subject to large amounts of foreground extinction, even in the near-IR, and this needs to be properly accounted for in order for accurate distances to be derived.

While dust maps obtained with tracers such as red clump and red giant stars are available \citep{Gonzalez-2012,surat2020}, a comparison between the observed and expected absolute magnitudes of Miras in different passbands also provides a route toward measuring foreground extinction {\em on a star-by-star basis}. Thus, for instance, following \citet[][]{Matsunaga-2009}, extinction in the $K_s$ band can be obtained from the observed $J-K_s$ or $H-K_s$ colors using the following expressions: 

  \begin{equation}
     A_{K_{s}}^{JK_{s}} = \frac{\left(\langle J\rangle -\langle K_{s}\rangle\right)-\left(M_J-M_{K_{s}}\right)}{r_J - 1}, \label{eq:aks-jk}
  \end{equation}

  \begin{equation}
     A_{K_{s}}^{HK_{s}} = \frac{\left(\langle H\rangle -\langle K_{s}\rangle\right)-\left(M_H-M_{K_{s}}\right)}{r_H - 1},  \label{eq:aks-hk}
  \end{equation}

\noindent where $r_J \equiv A_J / A_{K_{s}}$ and $r_H \equiv A_H / A_{K_{s}}$. 

Both $r_J$ and $r_H$ in these expressions depend on the adopted extinction law. The extinction law toward the Galactic bulge is complex and nonstandard, and several authors have provided their own calibrations. Distances derived using Eqs.~\ref{eq:aks-jk} and \ref{eq:aks-hk} will naturally depend on the adopted extinction law, and the differences, along with calibration issues such as those detected by \citet{Hajdu2020} in the case of VVV (which were corrected for in this study as described in Sect.~\ref{sec:calib}), may affect the astrophysical interpretation of the results \citep[e.g.,][]{Matsunaga-2016, Dekany-2019}. In this work, we have performed calculations using different prescriptions, including a (standard) \citet{Cardelli-1989} law. For ease of reference, the $r_J$ and $r_H$ values corresponding to the different laws examined in this paper are summarized in Table~\ref{tab:extlaw}. 

\begin{figure}
   \centering
   \includegraphics[width=0.45\textwidth]{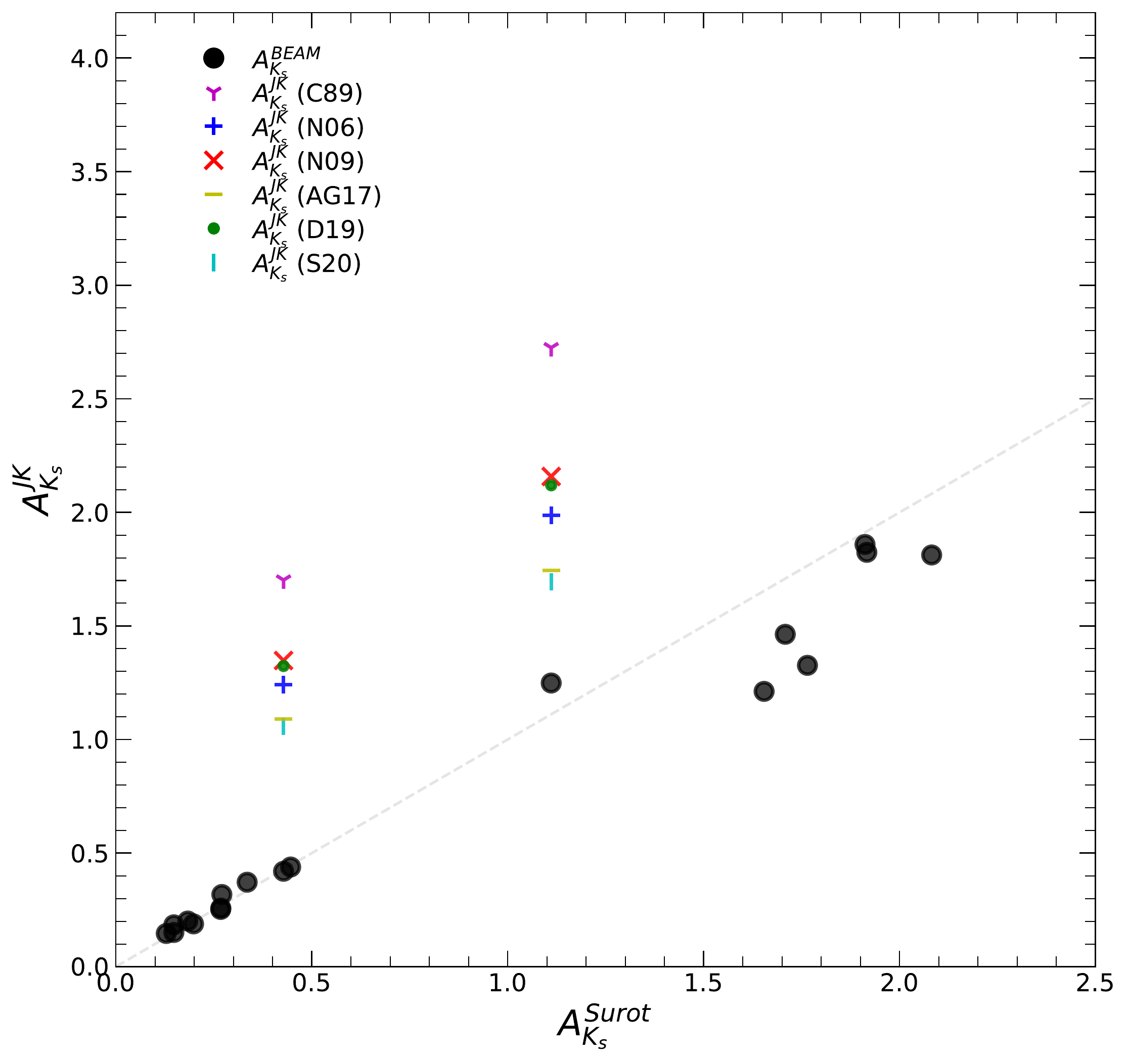}
   \includegraphics[width=0.45\textwidth]{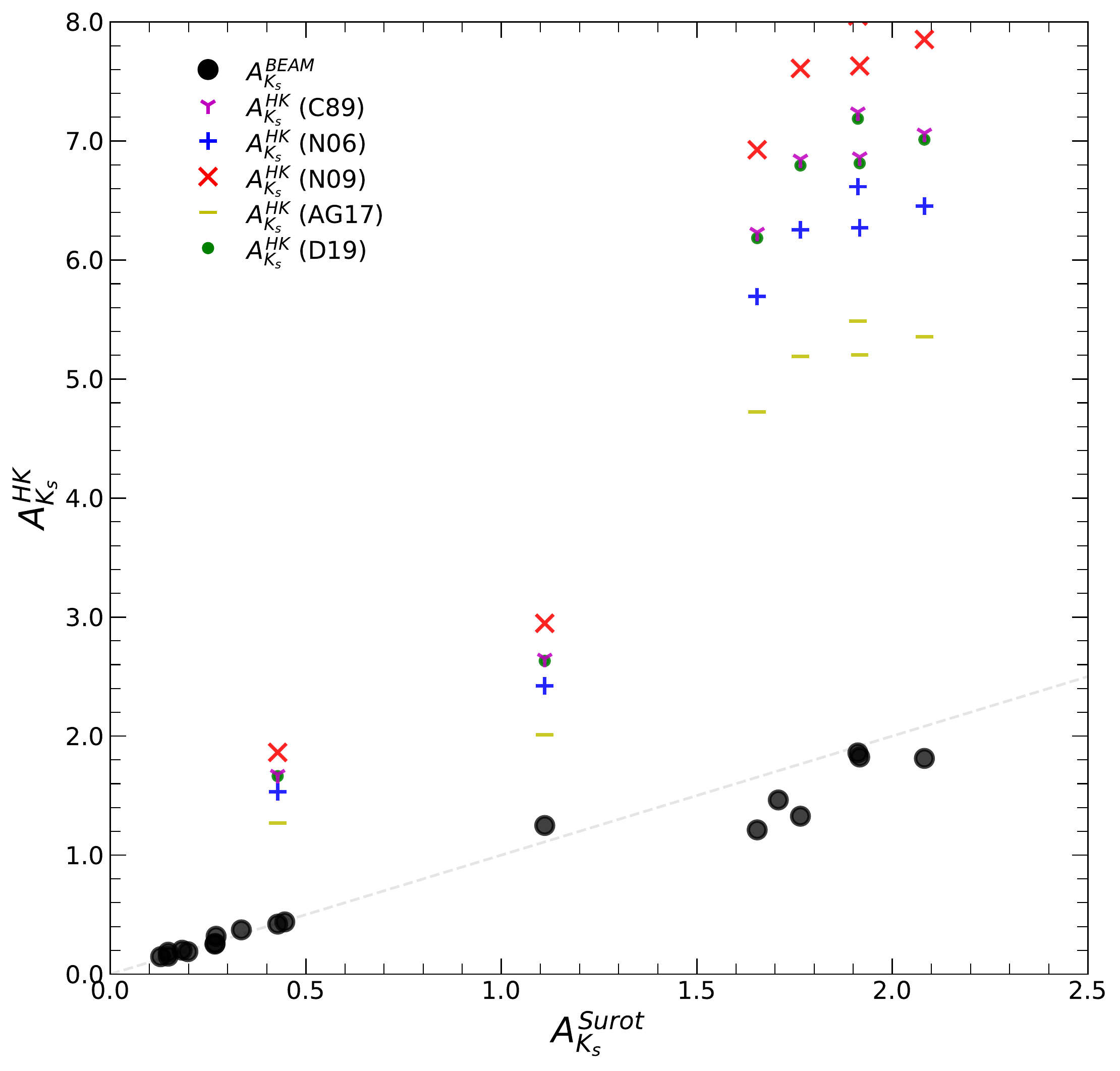}
              \caption{For nonsaturated VVV-PSF stars, the intrinsic extinction $A_{K_{s}}^{JK_{s}}$ (top panel) and $A_{K_{s}}^{HK_{s}}$ (bottom panel) is shown as a function of $A_{K_{s}}^{\rm Surot}$, according to the indicated extinction laws (see Table~\ref{tab:extlaw}). The same $A_{K_{s}}^{\rm BEAM}$ values are shown in both panels, obtained from $E(J-K_s)$ reddenings using a \citet{surat2020} extinction law, which is also used to obtain $A_{K_{s}}^{\rm Surot}$ values. The dashed gray line indicates a perfect correlation.}
              \label{fig:reddening}
\end{figure}

\begin{table}
\caption{Extinction Laws in this Work \tablefootmark{(a)} }
\label{tab:extlaw}      
\centering  
\begin{tabular}{l c c}
\hline
\hline
  \multicolumn{1}{c}{Law} &
  \multicolumn{1}{c}{$r_J$} & 
  \multicolumn{1}{c}{$r_H$} \\
\hline
\hline
  C89    &   2.47   &   1.67  \\ 
  N06    &   3.02   &   1.73  \\
  N09    &   2.86   &   1.60  \\
  AG17   &   3.30   &   1.88  \\
  D19    &   2.89   &   1.67 \\
  S20    &   3.37   &         \\
\hline
\hline
\end{tabular}
\tablefoot{
\tablefoottext{a}{
C89~= \citet{Cardelli-1989};
N06~= \citet{Nishiyama-2006}; 
N09~= \citet{Nishiyama-2009};  
AG17~= \citet{Alonso-Garcia-2017}; 
D19~= \citet{Dekany-2019}; 
S20~= \citet{surat2020}.
}
}
\end{table}

\begin{figure} [!ht]
   \centering
   \includegraphics[width=0.45\textwidth]{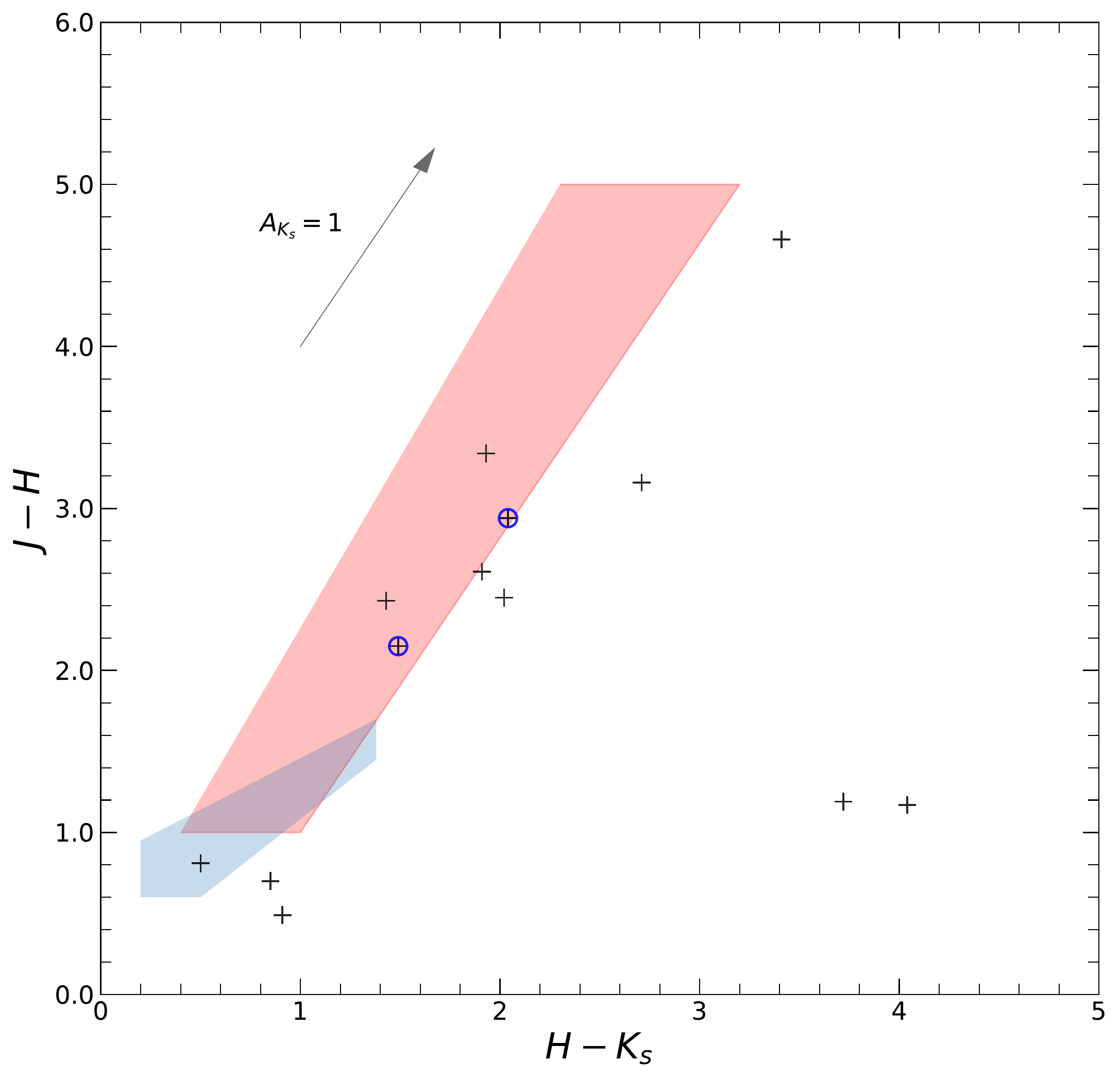}
   \caption{ Color-color ($J-H$ vs. $H-K_{s}$) plot for the nonsaturated candidate Miras in our sample for which measurements in the three bandpasses could be obtained (crosses). The two stars in our distance indicator sample are marked with blue circles. The shaded blue and red regions are schematic depictions of the loci occupied by Miras and SRVs in the \citet{Kerschbaum-2001} and \citet{Matsunaga-2009} samples, respectively. The arrow at the top of the figure indicates the reddening vector for an absolute extinction of 1~mag in $K_s$, assuming a D19 extinction law (see Table~\ref{tab:extlaw}).} 
   \label{fig:color-color}
\end{figure}

In our work, when deriving distances and extinctions, intensity-mean magnitudes in $J$, $H$, and $K_s$ from our VVV-PSF photometry are used when available. Absolute magnitudes are in turn obtained from the periods using the Mira period-luminosity relations (Eqs.~\ref{eq:plj}-\ref{eq:plks}). We call the thus derived, star-by-star extinction values {\em intrinsic extinctions} in what follows. 

In addition, we used both the Bulge Extinction and Metallicity Calculator\footnote{\text{\url{https://www.oagonzalez.net/beam-calculator}}} \citep[BEAM;][]{Gonzalez-2012}
and the latest \citet[][]{surat2020} extinction maps\footnote{\url{http://basti-iac.oa-teramo.inaf.it/vvvexmap/}} to obtain $E(J-K_{s})$ reddening values, based on which extinction values $A_{K_{s}}^{\rm BEAM}$ and $A_{K_{s}}^{\rm Surot}$ were obtained, assuming an extinction law as proposed by \citet{surat2020}.
We note that the recent Surot maps have higher spatial resolution than provided by BEAM. The latter is designed to provide extinction values at spatial resolutions not better than $2\arcmin$. Therefore, $A_{K_{s}}^{JK_{s}}$ and $A_{K_{s}}^{HK_{s}}$ should in principle provide better estimations, if reliable VVV-PSF photometry is available, as $A_{K_{s}}^{JK_{s}}$ and $A_{K_{s}}^{HK_{s}}$ are both obtained on a star-by-star basis. The Surot maps, on the other hand, have sub-arcmin spatial resolution, and thus its extinction values should be more directly comparable to those that we derive. 

The final extinctions for our distance indicator Miras are provided in Table~\ref{table:extinctions}, whereas Fig.~\ref{fig:reddening} shows how the values computed using different extinction laws and reddening maps compare with one another other. There is a clear tendency for our color-based values to be higher than their respective BEAM and Surot ones. This can plausibly be ascribed to depth effects, as the BEAM and Surot maps are two-dimensional and based primarily on red clump bulge stars, thus being incapable of tracing any dust that may be present on the far side of the bulge and beyond, where most of our candidates appear to be located (see below).

\begin{table*}
\caption{Extinction Values for Distance Indicator Miras\tablefootmark{(a)}}  
\scriptsize 
\label{table:4dis}
\label{table:extinctions}      
\centering 
\begin{tabular}{l c c c c c c c c c c c c c c}
\hline
\hline
  \multicolumn{1}{c}{ID\tablefootmark{(b)}} &
  \multicolumn{1}{c}{$A_{K_{s}}^{\rm BEAM}$} &
  \multicolumn{1}{c}{$A_{K_{s}}^{\rm Surot}$} &
  \multicolumn{6}{c}{$A_{K_{s}}^{JK_{s}}$} &&
  \multicolumn{5}{c}{$A_{K_{s}}^{HK_{s}}$} \\ 
  \cline{4-9} \cline{11-15} \\
  \multicolumn{1}{c}{} &
  \multicolumn{1}{c}{} &
  \multicolumn{1}{c}{} &
  \multicolumn{1}{c}{(C89)} &
  \multicolumn{1}{c}{(N06)} & 
  \multicolumn{1}{c}{(N09)} & 
  \multicolumn{1}{c}{(AG17)} &
  \multicolumn{1}{c}{(D19)} & 
  \multicolumn{1}{c}{(S20)} && 
  \multicolumn{1}{c}{(C89)} &
  \multicolumn{1}{c}{(N06)} & 
  \multicolumn{1}{c}{(N09)} & 
  \multicolumn{1}{c}{(AG17)} & 
  \multicolumn{1}{c}{(D19)} \\
\hline
\hline
302108018024 & 0.2 & 0.18 &   &   &   &   &   &   &&   &   &   &   &   \\      
305410041654 & 0.26 & 0.27 &   &   &   &   &   &   &&   &   &   &   &   \\   
307102042261 & 0.18 & 0.15 &   &   &   &   &   &   &&   &   &   &   &   \\     
307304149161 & 0.15 & 0.13 &   &   &   &   &   &   &&   &   &   &   &   \\     
307408087155 & 0.15 & 0.15 &   &   &   &   &   &   &&   &   &   &   &   \\       
308616154874 & 0.19 & 0.20 &   &   &   &   &   &   &&   &   &   &   &   \\     
309203069542 & 0.32 & 0.27 &   &   &   &   &   &   &&   &   &   &   &   \\    
309209099809 & 0.42 & 0.43 & 1.70 & 1.24 & 1.35 & 1.09 & 1.32 & 1.06 && 1.68 & 1.53 & 1.86 & 1.27 & 1.66 \\ 
309304135416 & 0.25 & 0.27 &   &   &   &   &   &   &&   &   &   &   &   \\       
319213010190 & 1.46 & 1.71 &   &   &   &   &   &   &&   &   &   &   &   \\       
319302072957 & 1.82 & 1.92 &   &   &   &   &   &   && 6.86 & 6.27 & 7.63 & 5.2 & 6.81 \\  
319509072340 & 1.33 & 1.77 &   &   &   &   &   &   && 6.85 & 6.25 & 7.61 & 5.19 & 6.79 \\ 
320106023807 & 1.25 & 1.11 & 2.72 & 1.99 & 2.16 & 1.75 & 2.12 & 1.69 && 2.65 & 2.42 & 2.95 & 2.01 & 2.63 \\ 
320201006873 & 1.86 & 1.91 &   &   &   &   &   &   && 7.24 & 6.62 & 8.05 & 5.49 & 7.19 \\
320201014922 & 1.81 & 2.08 &   &   &   &   &   &   && 7.06 & 6.45 & 7.85 & 5.35 & 7.01 \\  
320201073554 & 1.21 & 1.65 &   &   &   &   &   &   && 6.23 & 5.69 & 6.93 & 4.72 & 6.18 \\  
320515091949 & 0.44 & 0.45 &   &   &   &   &   &   &&   &   &   &   &   \\     
364104082076 & 0.37 & 0.34 &   &   &   &   &   &   &&   &   &   &   &   \\      
\hline
\hline
\end{tabular}
\tablefoot{
\tablefoottext{a}{For each nonsaturated star in our sample of distance indicator Miras (whose IDs are shown in column~1), extinction values are given according to the BEAM and Surot maps (columns 2 and 3, respectively). Columns 4 through 9 show the extinction values obtained using the stars' individual $(J-K_s)$ colors, according to the indicated extinction laws (see Table~\ref{tab:extlaw}). Similarly, columns 10 through 14 indicate extinction values obtained on the basis of the same extinction laws, except for S20 which is not available in the $H$ band. We note that, as explained in the text, the extinction values derived on the basis of the BEAM and Surot reddening maps should be considered as lower limits only, in the case of stars located beyond the Galactic bulge.}}
\end{table*}

\begin{table*}
\caption{Distances (in kpc) for Distance Indicator Miras\tablefootmark{(a)}}  
\scriptsize 
\label{table:13dis}
\label{table:distances}      
\centering 
\begin{tabular}{l c c c c c c c c c c c c c c}
\hline
\hline
  \multicolumn{1}{c}{ID\tablefootmark{(b)}} &
  \multicolumn{1}{c}{$D^{\rm BEAM}$} &
  \multicolumn{1}{c}{$D^{\rm Surot}$} &
  \multicolumn{6}{c}{$D^{JK_{s}}$} &&
  \multicolumn{5}{c}{$D^{HK_{s}}$} \\ 
  \cline{4-9} \cline{11-15} \\
  \multicolumn{1}{c}{} &
  \multicolumn{1}{c}{} &
  \multicolumn{1}{c}{} &
  \multicolumn{1}{c}{(C89)} &
  \multicolumn{1}{c}{(N06)} & 
  \multicolumn{1}{c}{(N09)} & 
  \multicolumn{1}{c}{(AG17)} &
  \multicolumn{1}{c}{(D19)} & 
  \multicolumn{1}{c}{(S20)} && 
  \multicolumn{1}{c}{(C89)} &
  \multicolumn{1}{c}{(N06)} & 
  \multicolumn{1}{c}{(N09)} & 
  \multicolumn{1}{c}{(AG17)} & 
  \multicolumn{1}{c}{(D19)} \\
\hline
\hline
302108018024 & 152.8 & 154.0 & &   &   &   &   &   &&   &   &   &   &   \\
305410041654 & 586.8 & 584.1 &   &   &   &   &   &   &&   &   &   &   &   \\
307102042261 & 253.7 & 258.0 &   &   &   &   &   &   &&   &   &   &   &   \\
307304149161 & 198.8 & 200.4 &   &   &   &   &   &   &&   &   &   &   &   \\
307408087155 & 144.0 & 144.2 &   &   &   &   &   &   &&   &   &   &   &   \\
308616154874 & 344.6 & 343.0 &   &   &   &   &   &   &&   &   &   &   &   \\
309203069542 & 150.3 & 153.6 &   &   &   &   &   &   &&   &   &   &   &   \\
309209099809 & 140.4 & 139.9 & 77.8 & 96.2 & 91.6 & 103.2 & 92.6 & 104.7 && 78.8 & 84.2 & 72.3 & 94.9 & 79.2 \\
309304135416 & 206.3 & 204.8 &   &   &   &   &   &   &&   &   &   &   &   \\
319213010190 & 132.9 & 118.7 &   &   &   &   &   &   &&   &   &   &   &   \\
319302072957 & 53.2 & 51.0 &   &   &   &   &   &   && 5.2 & 6.9 & 3.7 & 11.2 & 5.4 \\
319509072340 & 59.1 & 48.3 &   &   &   &   &   &   && 4.7 & 6.1 & 3.3 & 10.0 & 4.8 \\
320106023807 & 70.8 & 75.5 & 35.9 & 50.4 & 46.6 & 56.3 & 47.4 & 57.7 && 37.1 & 41.2 & 32.4 & 49.9 & 37.5 \\
320201006873 & 50.8 & 49.6 &   &   &   &   &   &   && 4.3 & 5.7 & 2.9 & 9.6 & 4.4 \\ 
320201014922 & 51.9 & 45.9 &   &   &   &   &   &   && 4.6 & 6.1 & 3.2 & 10.2 & 4.7 \\
320201073554 & 108.4 & 88.4 &   &   &   &   &   &   && 10.7 & 13.8 & 7.8 & 21.5 & 11.0 \\
320515091949 & 175.9 & 175.3 &   &   &   &   &   &   &&   &   &   &   &   \\
364104082076 & 162.3 & 165.1 &   &   &   &   &   &   &&   &   &   &   &   \\
\hline
\hline
\end{tabular}
\tablefoot{
\tablefoottext{a}{For each nonsaturated star in our sample of distance indicator Miras (whose IDs are shown in column~1), distance values (in kpc) are given on the basis of the apparent distance moduli in $K_s$, corrected (in each case) according to the extinction values given in Table~\ref{table:extinctions}. We note that, as explained in the text, the distances derived on the basis of the BEAM and Surot extinction maps, for stars located beyond the Galactic bulge, should be considered as upper limits only.}}
\end{table*}

True distance moduli are computed as $(m-M)_0 = (\langle K_{s}\rangle - A_{K_{s}}) - M_{K_{s}}$, using the different prescriptions for $A_{K_{s}}$ that were just described, whereupon distances $D$ (in pc) were finally obtained using $(m-M)_0 = 5 (\log D - 1)$. The latter distances are shown in Table~\ref{table:distances}. In that table, we refer to distances derived using $A_{K_{s}}^{JK_{s}}$ as $D^{JK_{s}}$, whereas those obtained using $A_{K_{s}}^{HK_{s}}$ are termed $D^{HK_{s}}$ instead. Also, $D^{\rm BEAM}$ and $D^{\rm Surot}$ are the distances obtained adopting extinctions based on the BEAM and Surot maps, respectively, in the manner described in the previous paragraph. 

\begin{figure} [!ht]
   \centering
   \includegraphics[width=0.45\textwidth]{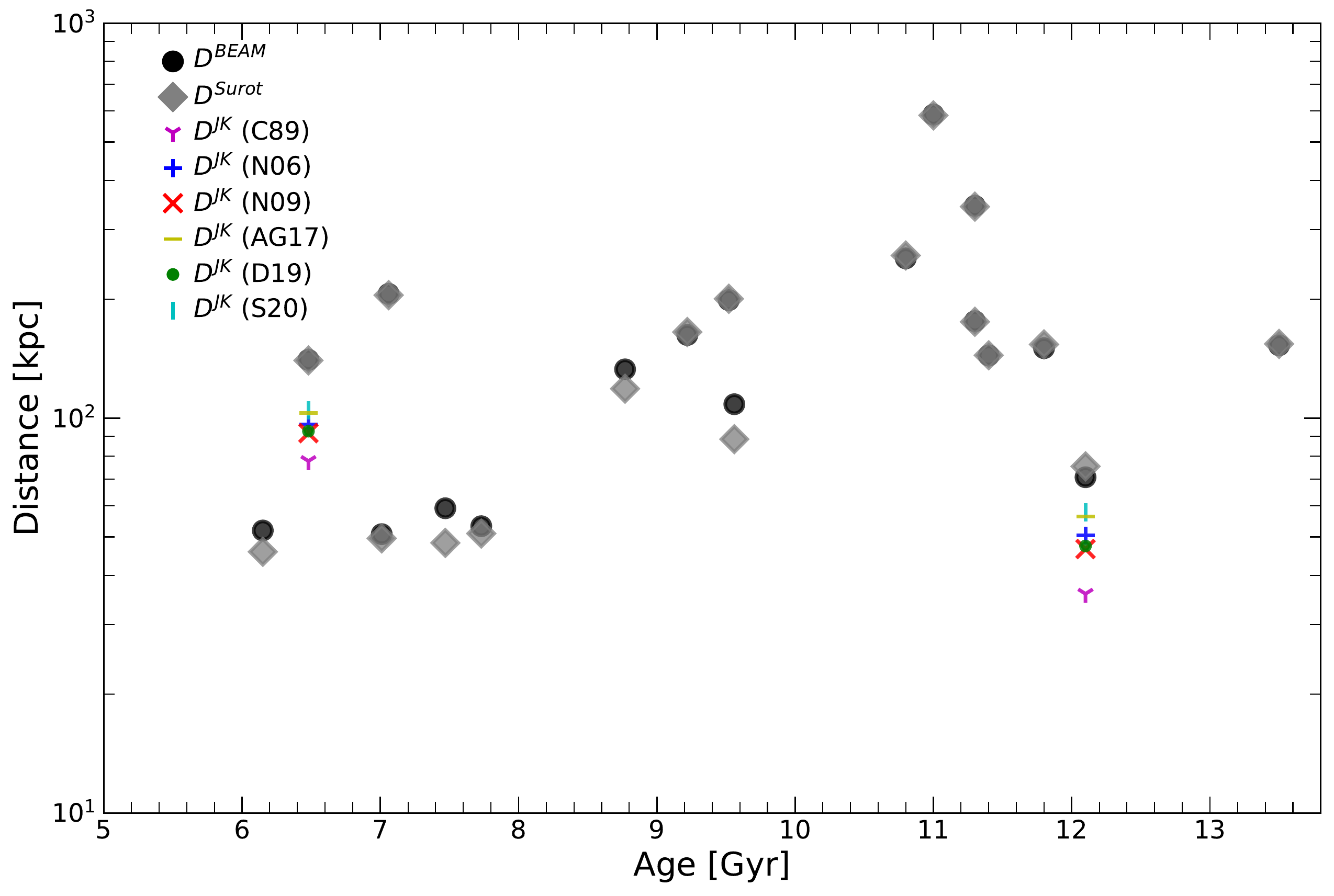}
   \includegraphics[width=0.45\textwidth]{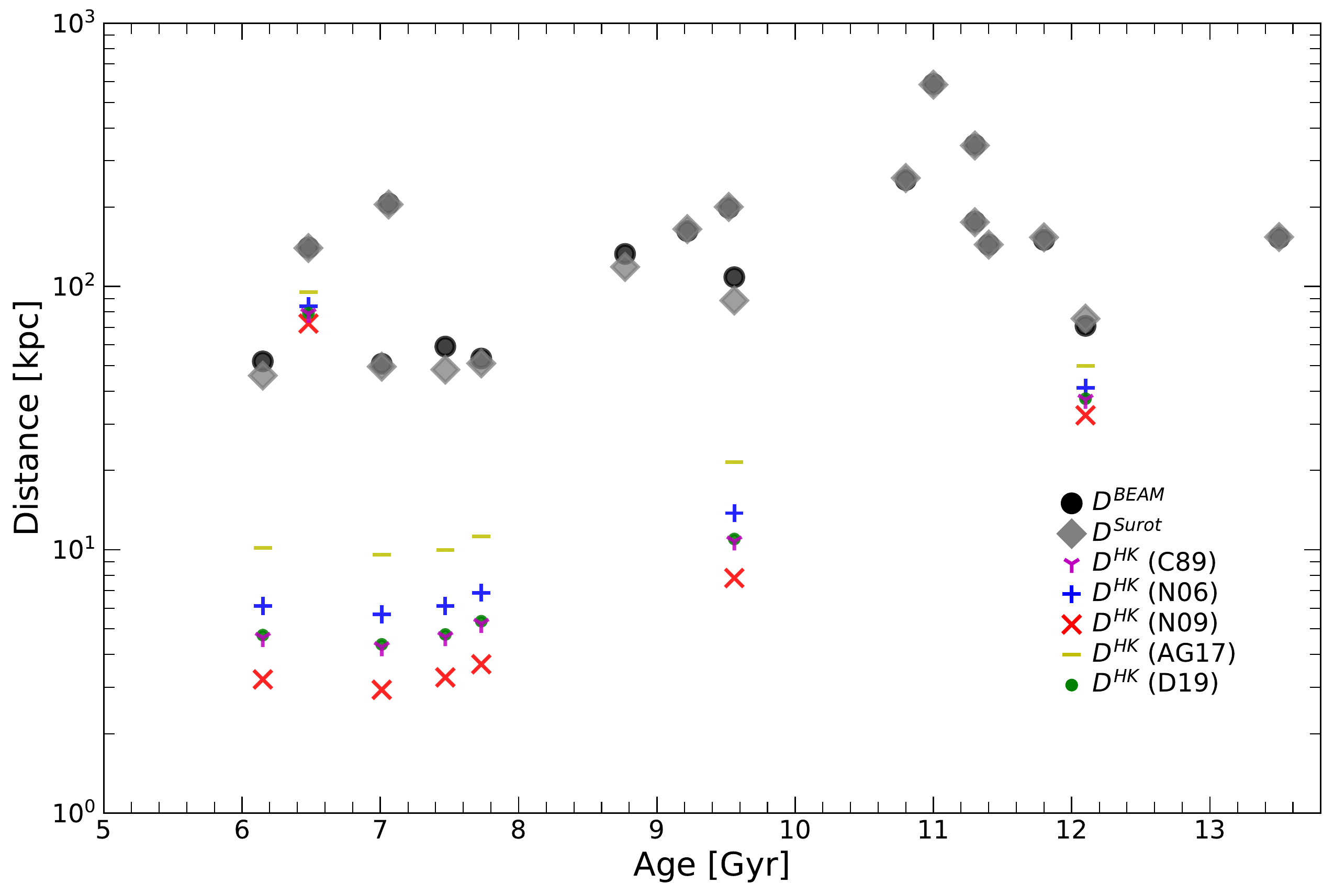}
   \caption{Distribution of distances (in kpc) versus ages (in Gyr) for the distance indicator Mira stars in our sample. $D^{JK_{s}}$ (top panel) and $D^{HK_{s}}$ (bottom panel) values are computed using the different indicated extinction laws (see Table~\ref{tab:extlaw}). 
   $D^{\rm BEAM}$ and $D^{\rm Surot}$ values are shown in both panels, and were computed from the $E(J-K_{s})$ values provided by the BEAM and Surot maps assuming a \citet{surat2020} extinction law. These maps are based primarily on bulge red clump stars, hence are not expected to accurately measure the reddening toward objects located beyond the Galactic bulge; accordingly, the corresponding distances should be considered as upper limits only.
   Due to lack of $J$-band measurements, several of the stars that are shown in the bottom panel with their respective $D^{HK_{s}}$ values do not have $D^{JK_{s}}$ data shown in the top panel (see also Table~\ref{table:distances}. Ages were computed following Eq.~\ref{eq:page-new}.  
   We note that the age range is restricted in this plot, as only Miras that have periods in the range $[100, 350]$~d (i.e., distance indicators) are included, corresponding to an age range of $[5.4, 13.6]$~Gyr according to Eq.~\ref{eq:page-new}.} 
   \label{fig:Dis vs. Age}
\end{figure}

The resulting $D^{JK_{s}}$ and $D^{HK_{s}}$ distance distributions of the VVV-PSF distance indicator Miras are shown in Figure~\ref{fig:Dis vs. Age}, plotted as a function of the Miras' ages (see Sect.~\ref{sec:ages}). As can be seen from this figure and Table~\ref{table:distances}, most of our distance-indicator Miras appear to be very distant, though their exact distances remain uncertain, with significant differences found for some stars based on the method used and extinction law assumed. As a rule, however, our $D^{JK_{s}}$ and $D^{HK_{s}}$ distances place them much closer than implied by $D^{\rm BEAM}$ or $D^{\rm Surot}$. This too can be understood in terms of depth effects, to which our color-based method is sensitive but which are not properly described by the BEAM and Surot maps. The end result is that the BEAM and Surot extinctions are likely underestimated, and the corresponding distances likely overestimated, compared with those computed using the stars' colors. We note, in particular, and in accordance with Table~\ref{table:extinctions} and Fig.~\ref{fig:reddening}, that a significant fraction of our stars may be subject to quite extreme levels of extinction, at a level that is not captured by either the BEAM or Surot maps. Such high extinction levels are consistent with the fact that some of these sources could not be reliably detected in the $J$ band. One possible problem with this explanation is the fact that the four stars that appear to be closest to us according to their $D^{HK_{s}}$ values (i.e., those with VVV IDs 319302072957, 319509072340, 320201006873, and 320201014922), are also the ones with the highest $A_{K_{s}}^{HK_{s}}$ extinction values (see Tables~\ref{table:extinctions} and \ref{table:distances}), which would appear to be contradictory, particularly since their corresponding $A_{K_{s}}^{\rm BEAM}$ and $A_{K_{s}}^{\rm Surot}$ values, which measure the integrated dust extinction up to the distance of the Milky Way bulge, are significantly smaller. Note, in addition, that there is one extinction law, namely AG17 (see Table~\ref{tab:extlaw}), that places all these stars sufficiently far away that they could still be on the far side of the bulge. Other possible explanations may include poor measurements (these stars are all quite faint in $H$, and indeed not detected in $J$) or their being C-rich stars. Further analysis will be required to properly understand their properties.

As a further check of the colors of the stars in our sample, in Figure~\ref{fig:color-color} we show the color-color distribution of the sources for which we could obtain both $J-H$ and $H-K{s}$ measurements, compared with the loci occupied by the Mira and SRV sources previously studied by \citet{Kerschbaum-2001} and \citet{Matsunaga-2009} (light blue and red polygons, respectively). As we can see from this plot, the majority of the stars in our samples for which colors could be measured fall reasonably close to the loci defined by the distributions of Miras and SRVs studied by these authors. However, as in \citet{Matsunaga-2009}, there is a significant fraction of stars in our sample that fall to the right of the main distribution, i.e., toward redder $H-K_{s}$ colors. Some could be symbiotic Miras \citep{Whitelock-1987,Whitelock-2003b}. On the other hand, uncertainties in the zero-point corrections that were introduced in Sect.~\ref{sec:calib} to alleviate the calibration issues that were pointed out by \citet{Hajdu2020} could affect the derived colors, particularly in the higher-density regions, as could the fact that for many stars the number of available $J$ and $H$ measurements is extremely small. However, there are several other possible explanations for these anomalous colors. As pointed out by \citet{Matsunaga-2009}, unresolved blends, some of which may be difficult to detect even upon close examination of the images, may affect the measured colors, by moving them either to the blue (if the contaminant is a foreground blue source) or to the red (if the contaminant is redder than the candidate Mira or SRV). In addition, we have verified that some of these stars fall within the regime of saturation in the VVV images, so that, as previously pointed out (Sect.~\ref{sec:LPVsel}), their magnitudes and colors must be treated with extreme caution. Also, for a few of them the available $J$ and/or $H$ measurements were all obtained close to the maximum of the light curve, thus rendering their corresponding mean magnitudes particularly uncertain: in fact, this is exactly what happens in the case of the two outliers in the bottom right of the diagram, whose VVV IDs are 320211006775 and 364207072166. Lastly, some stars were detected close to the limit of the VVV-PSF photometry, and thus their measurements can be quite unreliable. In particular, we have checked that these are indeed suitable explanations for the three sources that deviate the most to the red in the color-color diagram, at around $H-K_s \approx 4$, and the one that has a $J-H$ color close to zero. Further analysis of all of these sources with peculiar colors would be very valuable to place their near-IR magnitudes and colors, and indeed their very classification as Mira candidates, on a firmer footing.

Be as it may, our sample clearly differs from the one previously studied by \citet{Matsunaga-2009}, which presents a sharp peak around 8.2~kpc, which is essentially the distance to the Galactic center \citep[][and references therein]{Bland-Hawthorn-2016,Gravity-2019}. The VVV-PSF Miras for which distances could be reliably computed extend much farther out, with many stars seemingly located well beyond the Galactic bulge. This can be understood if we recall not only that the VISTA telescope that was used in the VVV survey has a 4-meter aperture, as compared to IRSF's 1.4-meter aperture, but also that the exposure times adopted in the VVV survey were sufficiently long that unprecedented depth could be reached. Conversely, the more nearby Miras are missing in our catalog because they are saturated in the VVV-PSF images, hence reliable distances could not be computed for them. Lastly, the volume surveyed on the far side of the bulge is vastly greater than that on its near side.

{\em Gaia} parallaxes can in principle also be used to obtain distances to LPVs. However, such stars may often not be point sources for the {\em Gaia} survey, so that their corresponding {\em Gaia} distances should be taken with due caution, considering that these are large stars that can have complex, structured photospheres, and angular diameters comparable to (or even larger than) their measured parallaxes \citep[][]{Whitelock-2012}. Keeping this caveat in mind, we have cross-matched our catalog against {\em Gaia}'s DR2 and early DR3 \citep[eDR3;][]{GaiaeDR3-2021}. As pointed out in Sect.~\ref{sec:crossmatch}, most of the matches that we found correspond to saturated sources in our VVV-PSF photometry. There are, however, two exceptions, namely stars with VVV IDs 307408087155 and 309203069542. 
Both of these sources have periods that place them within the realm of distance indicators, and thus we were able to compute distances for all of them. Indeed, according to Table~\ref{table:distances}, these sources are located at distances of roughly 144~kpc and 150~kpc, respectively. These are rather extreme distances which, if correct, would make derivation of their parallaxes extremely challenging, even for {\em Gaia} \citep[see also][who find negative or insignificant parallaxes for their outer halo LPVs, whose distances they report to be in the range between 50 and 119~kpc]{Mauron-2021}. On the other hand, since the only available extinction values for these stars are the ones obtained on the basis of the BEAM and Surot maps, their distances could be significantly overestimated (see Fig.~\ref{fig:Dis vs. Age} and Table~\ref{table:distances}). Accordingly, we carefully checked {\em Gaia}'s eDR3 catalog, and found the following results: 

\begin{itemize} 
\item VVV 307408087155: There are actually two {\em Gaia}'s eDR3 sources within 1\arcsec\ of this star. The closest one, Gaia eDR3 4062695911426083840, has a parallax value of $1.37 \pm 0.31$~mas, thus placing it at a distance of only about 739~pc from us. It is also relatively bright, with a reported $G = 16.7$. This source, which also looks bright in VVV images, is offset from the position of VVV 307408087155 by about 0.6\arcsec. We deem it very unlikely that our source, which is so faint as to even go undetected in $J$ and $H$, could correspond to this bright, nearby star. There is another source in the vicinity that is present in {\em Gaia}'s eDR3, namely Gaia eDR3 4062695915664384256, that is 0.8\arcsec\ away from the position of the VVV source. This star has a reported $G = 19.3$, and no parallax measurements reported. We believe that it is sufficiently far away in the sky that it is also unlikely to correspond to our detected VVV source. 

\item VVV 309203069542: There is only one source matching the VVV coordinates within a radius or 1\arcsec, namely Gaia eDR3 4064127338032497024. At $G = 20.1$, this is a very faint source in {\em Gaia}'s eDR3, which assigns to it a negative parallax, thus making it impossible to compute a reliable astrometric distance. 

\end{itemize}

The majority of our distance-indicator Miras appear to be located well beyond the Galactic bulge: some of them may be associated to distant MW arms, background halo stars,  heretofore unknown globular clusters or dwarf spheroidal galaxies, or even the Sagittarius dwarf spheroidal galaxy \citep{Ibata-1994}, whose main body is located at a distance of about 26~kpc \citep{McConnachie-2012}, and whose tidal arms extend across much of the sky \citep[e.g.,][and references therein]{Navarrete-2017,Ramos-2020}.
 
We note that the distances of some of our stars place them far out in the MW halo, some being even close to (one even more distant than) the predicted halo edge, which \citet{Deason-2020} place at a distance of $292 \pm 61$~kpc. One should keep in mind, however, that for many of the stars with the largest distances, the latter were derived on the basis of BEAM and Surot extinctions, and are thus likely overestimated. On the other hand, using light curves from the Catalina Sky Surveys \citep{Drake-2017}, \citet{Mauron-2021} were similarly able to identify ten LPVs in the outer halo, with distances in the range $\sim 50-120$~kpc, and thus not reaching as far out into the outermost halo as we do in this study. On the other hand, \citet{Mauron-2021}
were able to use data from {\em Gaia}'s DR3 \citep{GaiaeDR3-2021} to show that the tangential velocities of their 10 distant LPVs are consistent with expectations based on other bona-fide halo objects, such as globular clusters and dwarf galaxies. 

Given the fact that the period range of our distance indicator Miras is capable of tracing intermediate- and old-age populations only (see Sect.~\ref{sec:ages}), even the closest among our stars with measured distances are unlikely members of spiral arms, but may more likely belong to the intermediate-age components of the MW thick disk \citep[e.g.,][]{Robin-2003,Giammaria-2021}, dwarf satellite galaxies, or even halo \citep[see also][]{Grady-2019,grady-2020}. Some of the closer stars, if their $D^{HK_{s}}$ distances are to be trusted, could in principle belong to the MW bulge, the latter being a complex system with a radius of $\sim 2.5$~kpc \citep[][and references therein]{Zoccali-Valenti-2016, Barbuy-2018, Zoccali-2019}, though this may perhaps seem unlikely from a statistical perspective, given the bulge's fairly restricted volume of space compared to that covered by our sample, and also the fact that the BEAM and Surot maps place them much farther away. On the other hand, BEAM- and Surot-based extinction values for individual stars, and particularly the more distant ones, are likely underestimated, since, as previously discussed, the BEAM and Surot maps do not count the extinction component that is due to dust lanes located beyond the Galactic bulge. If so, BEAM- and Surot-based distances for our more distant stars may have been correspondingly overestimated, and thus their true distances may potentially be much smaller, given the very significant differences between the Surot/BEAM extinction values and the ``intrinsic'' ones computed for a few of our stars (see Tables~\ref{table:extinctions} and \ref{table:distances}). 
On the other hand, color-based extinctions can also be inaccurate, due to insufficient and/or incorrect data and/or measurements in the $J$ and/or $H$ bands. Clearly, further analysis will be needed in order to place the distances of our Miras at a firmer footing.

To close, we note that VVV data have also afforded the detection of faraway structures beyond the MW bulge based on other tracers, such as Cepheids, red clump stars, and microlensing events \citep[e.g.,][]{Dekany-2015L,Dekany-2015,Minniti-2018ebv,Dekany-2019,Navarro-2020-ml,JMinniti-2020,Saito-2020}. 

\begin{figure} [!ht]
   \centering
   \includegraphics[width=0.45\textwidth]{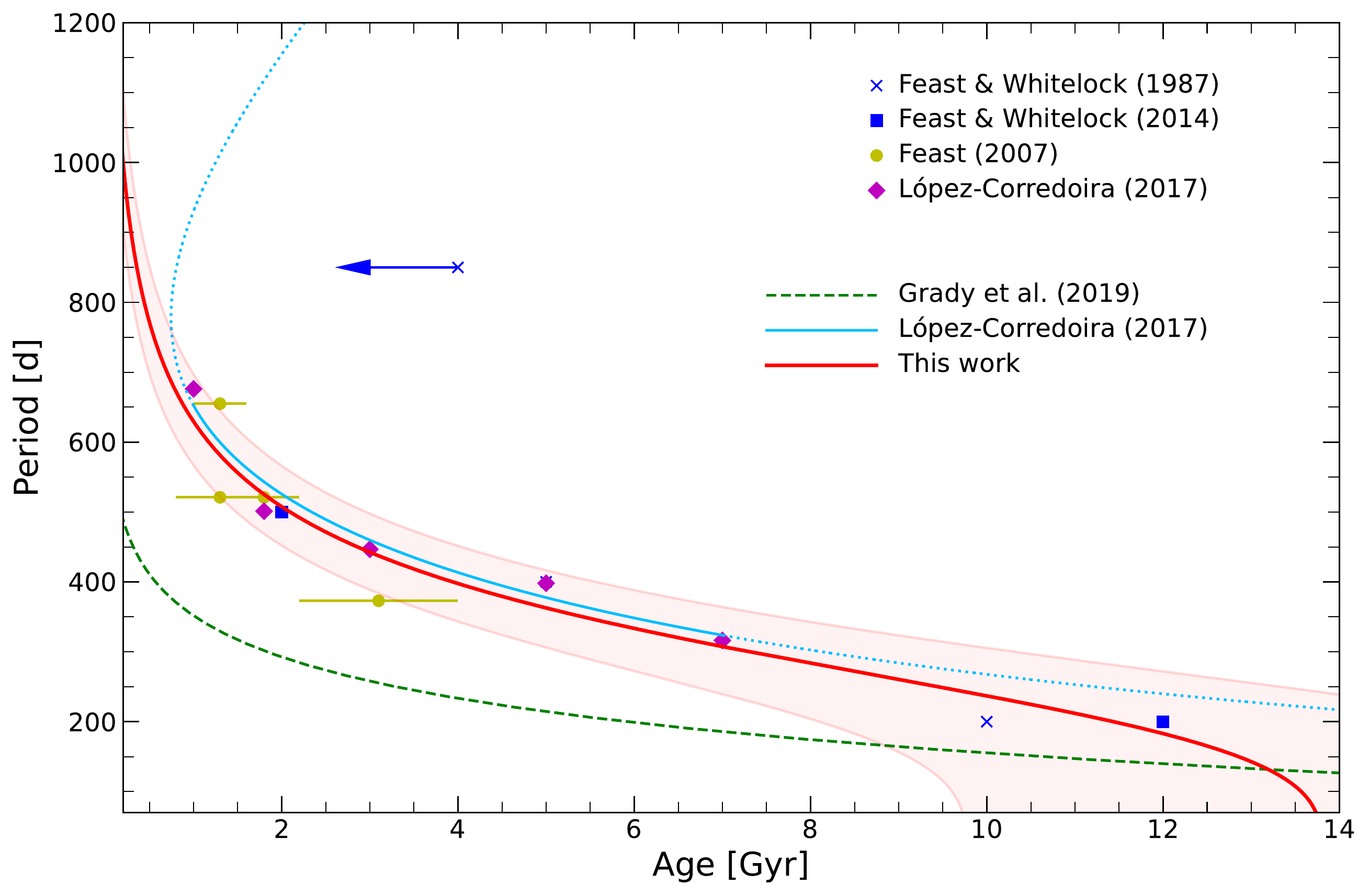}
   \includegraphics[width=0.45\textwidth]{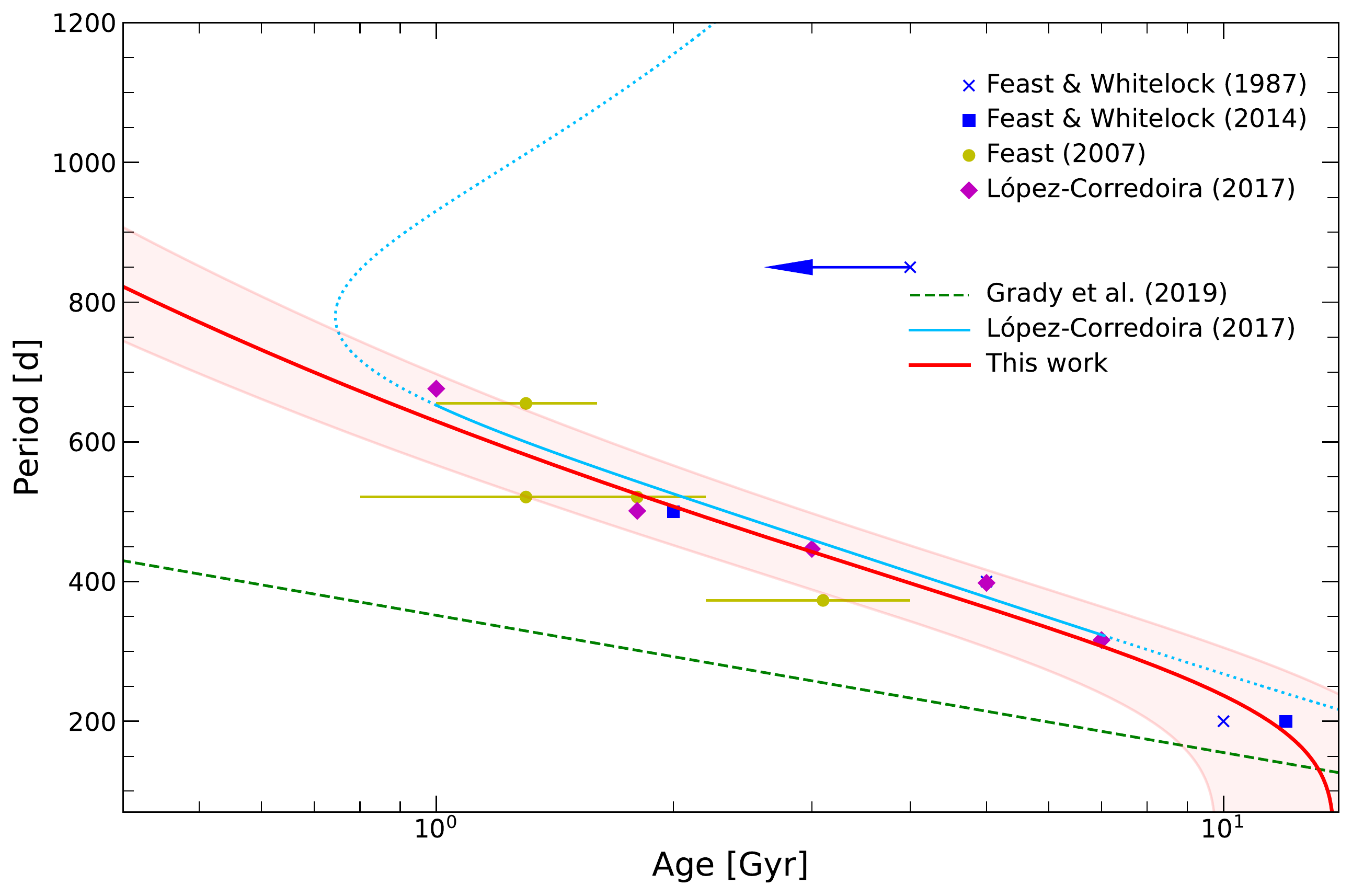}
   \caption{Period-age relations for Mira variables, shown in a linear scale (upper panel) and using a logarithmic scale for the ages (bottom panel). Blue crosses, blue squares, yellow circles, and magenta diamonds represent datapoints extracted from \citet{Feast-1987,Feast-Whitelock-2014}, \citet{Feast-2007}, and \citet{Lopez-Corredoira-2017}, respectively. The blue arrows indicate that only an upper limit on the age is available. The green dashed line represents the relation suggested by \citet{Grady-2019} whereas the light blue solid line the relation by \citet{Lopez-Corredoira-2017}. The light blue dotted line segments illustrate the expected behavior of the \citet{Lopez-Corredoira-2017} relation, when it is extrapolated beyond its formal range of validity. The red line is the relation proposed in this work (Eq.~\ref{eq:page-new}), and the light red band represents a dispersion of $\Delta \log t = \pm 0.15$ around it, which basically encompasses all available measurements in the literature, but may be an underestimate at the long-period/short-age end in particular, where empirical data are not yet available.} 
   \label{fig:page-relation}
\end{figure}

\subsection{Period-age relation}\label{sec:ages}

As pointed out in Sect.~\ref{sec:intro}, different prescriptions that are available in the literature for the relationship between periods and ages of Miras provide different results. This can be clearly seen from Figure~\ref{fig:page-relation}, where the analytical expressions provided by \citet[][]{Lopez-Corredoira-2017} and \citet[][]{Grady-2019} are compared, and data inferred from a variety of studies and compilations \citep{Feast-1987, Feast-Whitelock-2014, Feast-2007, Lopez-Corredoira-2017} are also overplotted. As can be clearly seen, the \citet[][]{Grady-2019} relation does not match the available data well. The \citet[][]{Lopez-Corredoira-2017} relation, on the other hand, provides a reasonable description of the data at the intermediate ages where his relation was calibrated, but its quadratic nature leads to unphysical results when extrapolated toward both the short and long period ends. To compensate for these shortcomings, here we propose the following expression:

\begin{equation} 
{\rm Age} = \frac{a}{1 + { \left(\frac{P}{c}\right)^b}},
\label{eq:page-new}
\end{equation} 

\noindent with $a = 13.8$~Gyr, $b = 3.6$, and $c = 310$~days. In this expression, the age is in Gyr, and the period $P$ is in days. As Figure~\ref{fig:page-relation} shows, this relation has the clear advantage of describing correctly the expected dependence of periods on ages at both the young and old age extremes, while at the same time providing a reasonable description of the intermediate-age regime as well. Note, in particular, that Eq.~\ref{eq:page-new} tends asymptotically to a value of 13.8~Gyr at the short-period end, which is the current best estimate for the age of the Universe (\citeauthor{Bennett-2013} \citeyear{Bennett-2013}; \citeauthor{Planck-2016} \citeyear{Planck-2016}; see also \citeauthor{Catelan-2018} \citeyear{Catelan-2018} and \citeauthor{Valcin-2020} \citeyear{Valcin-2020}, for recent reviews and additional references). 

Several previous authors have used the Mira period-age relation to study how the ages of Miras change as a function of position across the different Milky Way components, including halo, disk, bulge, and bar \citep[e.g.,][]{Catchpole-2016,Lopez-Corredoira-2017,Grady-2019,grady-2020}. 
We adopt our improved relation with a similar purpose in what follows. However, we do caution that more age measurements, particularly at the young and old regimes where data are currently lacking, and at different metallicities, would be of great importance to place it on a firmer footing, as well as to constrain its possible dependence on chemical composition. Indeed, as also shown in Figure~\ref{fig:page-relation}, the still very limited data that are available for calibration purposes imply an uncertainty of $\Delta \log t \sim 0.15$ in the ages inferred from this expression, which may actually be an underestimate, particularly in the long-period/short-age regime. Similarly, theoretical models indicate that individual LPVs may undergo significant period changes in the course of their evolution \citep[e.g.,][]{Marigo-2017, Trabucchi-2021}, which would imply the existence of an additional, empirically unconstrained, intrinsic spread around the values implied by Eq.~\ref{eq:page-new}. Figure~4 in \citet{Marigo-2017}, where evolutionary models and isochrones that include the thermally pulsing AGB phase were presented, shows that the period evolution, for a given pulsation mode, may be comprised of a relatively smooth (and small) long-term trend, with larger superimposed variations taking place on much shorter timescales. While the long-term component would likely dominate in a random sample of Miras of a given age, a larger sample and detailed Monte Carlo modeling would be required to properly evaluate the most likely period values and corresponding dispersions. Such modeling would be of significant interest, but is outside the scope of our paper.

It should also be noted that the limited data used to obtain this relation corresponds to disk stars in the solar neighborhood. Thus, it is unclear whether Eq.~\ref{eq:page-new} can be applied in the case of stars belonging to different populations and with different metallicities. Observations of Miras in globular clusters suggest that the metallicity has a large impact on the period, at least among low-mass Miras \citep{Frogel-1998,Feast-2000ASSL}. There are known Miras in moderately metal-rich globular clusters \citep{Sloan-2010} whose periods, according to Eq.~\ref{eq:page-new}, would imply ages of only a few Gyr, which is likely several Gyr younger than the actual ages of these clusters \citep[e.g.,][]{Leaman-2013,VandenBerg-2013,Cohen-2021}. It has been suggested that the presence of long-period, C-rich Miras in globular clusters could be ascribed to the mergers of two old stars \citep{Feast-2013, Matsunaga-2017}. Naturally, for any such stars that are merger products, whether in clusters or in the field, the age implied by Eq.~\ref{eq:page-new} would be an underestimate, compared with the ages of the original binary pairs that led to their formation. This has indeed been suggested as a possible explanation for the observed properties of C-rich stars in the Milky Way bulge, whose periods, when taken at face value, would imply that they have ages of only a few Gyr \citep{Matsunaga-2017}.

\begin{figure}
   \centering
   \includegraphics[width=0.45\textwidth]{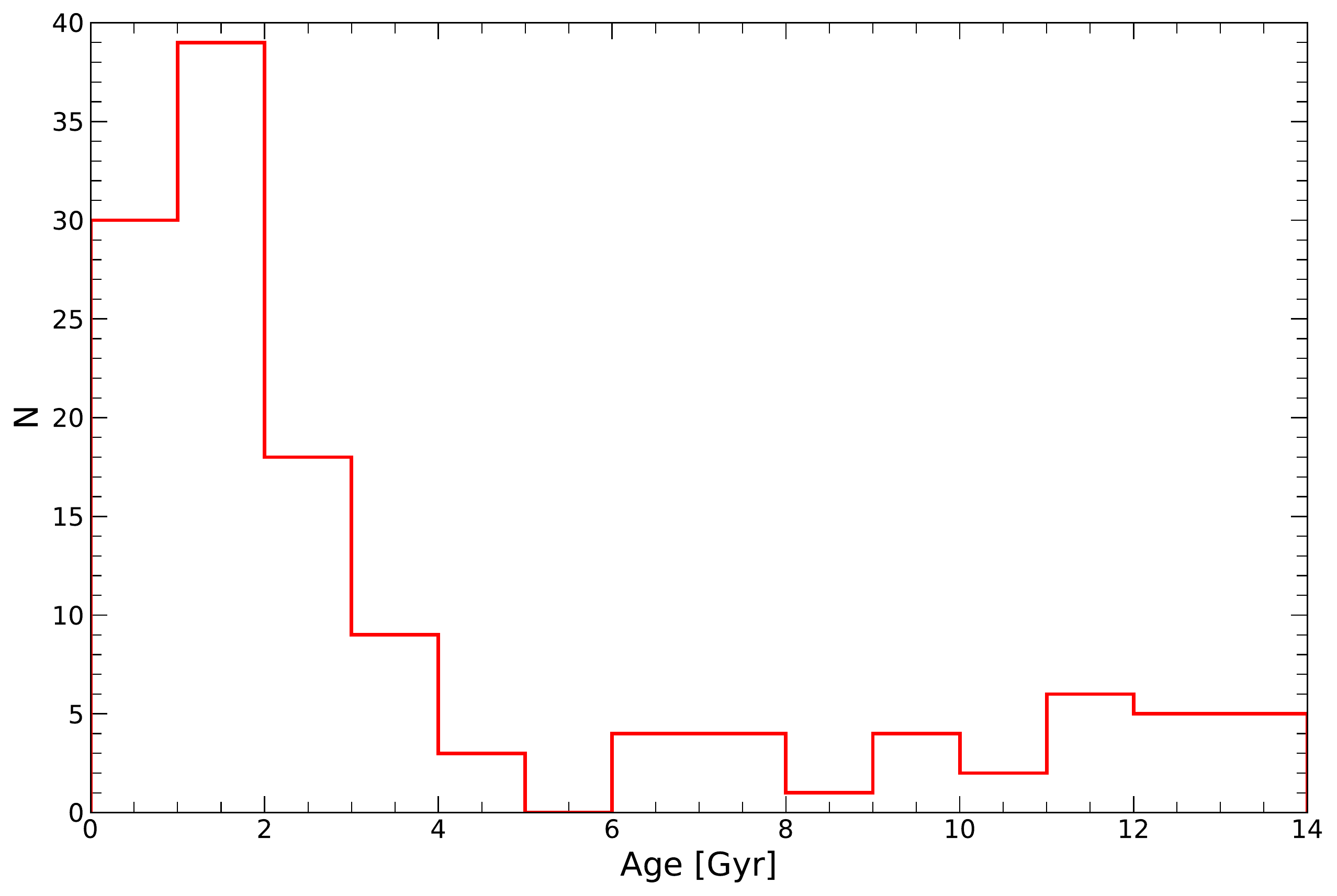}
   \caption{Histogram of ages for the Mira stars in our sample. The ages are given in Gyr, and were computed based on the stars' periods according to Eq.~\ref{eq:page-new}. The bimodal age distribution is a direct consequence of the bimodal distribution in periods (Fig.~\ref{Fig:his-p}): the minimum close to 365~d corresponds to an age of 4.9~Gyr, according to Eq.~\ref{eq:page-new}.} 
   \label{fig:his-age}
\end{figure}

A histogram showing the age distribution of our Mira candidates is presented in Figure~\ref{fig:his-age}. Bimodality is apparent in this plot, and is a direct consequence of the bimodal distribution in periods for the stars in our sample (Fig.~\ref{Fig:his-p}), the minimum close to 365~d corresponding to an age of 4.9~Gyr according to Eq.~\ref{eq:page-new}. 

The distribution of distances (listed in Table~\ref{table:13dis}) as a function of age is displayed in Figure~\ref{fig:Dis vs. Age} for our distance indicator Miras. $D^{JK_{s}}$ (top panel) and $D^{HK_{s}}$ (bottom panel) values are computed using the different indicated extinction laws (Table~\ref{tab:extlaw}), based on VVV-PSF magnitudes. In addition, $D^{\rm BEAM}$ and $D^{\rm Surot}$ values are shown in both panels. Since only distance indicator Miras are included in this plot, the age range is significantly limited, compared with Figure~\ref{fig:his-age}. Indeed, distance indicator Miras have periods in the range $[100, 350]$~d, which corresponds to an age range of $[5.4, 13.6]$~Gyr,  according to Eq.~\ref{eq:page-new}. 

In this very limited sample, and over this limited age range, there is no clear trend between distance and age, although a tendency of ages increasing with distance may be implied by the $D^{HK_{s}}$ values only. Indeed, there is significant dispersion in ages at any given distance. This could in principle imply that Miras belonging to different stellar populations are present in our dataset. On the other hand, and as the bottom plot of Figure~\ref{fig:Dis vs. Age} illustrates particularly well, any firm conclusions can only be derived once the uncertainty in the extinction values toward individual stars can be greatly reduced. Note, in this sense, that the seemingly reduced dispersion that is seen in the upper panel of Figure~\ref{fig:Dis vs. Age} is, at least in part, an artifact of the fact that $J$-band magnitudes are not available for several stars, whose $D^{JK_{s}}$ values are accordingly missing from that plot.

\section{Conclusions}\label{sec:concl}
Near-IR time-series data from the VVV survey were used to perform a systematic search for LPVs toward selected fields in the MW bulge. These fields were chosen in order to maximize overlap with the OGLE-III LPV catalog. Based on PSF photometry performed on the VVV images, we developed a new technique that uses both correlated and noncorrelated statistical indices in order to select Mira candidates, which was tested against the previously known OGLE stars in those fields and was found to produce excellent results. As a consequence, a sample of 130 Mira candidates was obtained. An additional 1013 LPV candidates are also reported (our LPV$^{+}$ candidates) for which the available data proved insufficient to obtain reliable periods, and thus provide a better assessment of their LPV status (Mira, SRV, or other). In both sets, and especially in the latter, other types of LPVs, in particular SRVs, may also be present. We provide a catalog with the full list of Miras and Mira candidates detected in this work. Our catalog includes coordinates, mean magnitudes, and amplitudes (obtained from a template-fitting procedure) in each of $J$, $H$, and $K_{s}$, periods, and ages. The latter were derived on the basis of a reassessment of previously proposed period-age relations in the literature, with a new analytical expression presented in this work. 

Among the stars in our catalog, a significant number of distance indicator Miras are present~-- i.e., Miras with periods in the range $[100, 350]$~d, over which the PL relation of Miras is known to be well behaved. Unfortunately, saturation affects the photometry of many of the stars in our sample, so that only a relatively small subset possesses reliable data that can be used for distance determination purposes. Conversely, extinction toward some of these stars can be so extreme, and their distances so large, that few or no reliable measurements in $J$ and/or $H$ exist. When the latter are available, however, extinction values were obtained on a star-by-star basis by comparing the intrinsic colors expected from the Mira PL relations in $J$, $H$, and $K_{s}$ with the actually observed colors (when available), and assuming different extinction laws that have been proposed in the literature. Distances were also obtained for the same stars using both the BEAM and Surot reddening maps, assuming a  \citet{surat2020} extinction law. We find that the distances obtained using reddening maps are systematically larger than those computed using our color-based method, which may plausibly be due to the fact that those maps fail to describe the presence of dust on the far side of the Milky Way bulge and beyond.

We find that the stars in our sample have a bimodal distribution in periods, implying also a bimodal distribution in ages, with the minimum located close to 365~d, corresponding to an age of approximately 4.9~Gyr. This result is likely a consequence of different sources of bias affecting the VVV data. For the small subsample of distance indicator Miras which are not saturated in the VVV data, we do not find clear evidence of an age-distance relation, obtaining instead a significant spread in ages at any given distance. Unlike previous studies that lacked the sensitivity and depth of VVV observations, we find a significant number of very distant Miras, some of which may belong to the halo or dwarf spheroidal galaxies (including Sagittarius). A few of our stars may even be located close to the edge of the Milky Way halo, although this conclusion hinges heavily on reddening maps that do not properly count the extinction caused by dust lanes on the other side of the Milky Way bulge, likely leading to overestimated distances. Accordingly, any definite conclusions regarding the spatial and age distribution of our distance indicator Miras will require a significant reduction in the uncertainty associated with extinction along the line of sight, including, most crucially, three-dimensional (i.e., distance-dependent) effects. 
In future papers of this series, we will apply the same techniques developed in this study to other MW fields covered by VVV.

\begin{acknowledgements}
We gratefully acknowledge the constructive feedback provided by the referees, whose comments and suggestions have led to a significantly improved manuscript.
We warmly thank Felipe Gran, Phil Lucas, Claudio A. Navarro Molina, and Zhen Guo for helpful discussions. Support for this project is provided by ANID's Millennium Science Initiative through grant ICN12\textunderscore 12009, awarded to the Millennium Institute of Astrophysics (MAS); by ANID's Basal projects AFB-170002 and FB210003; and by FONDECYT grant \#1171273. F.N is grateful for financial support by Proyecto Gemini CONICYT grants \#32130013 and \#32140036, and by P. Universidad Cat\'olica de Chile's Vicerrector\'{i}a de Investigaci\'on (VRI). C.E.F.L. acknowledges PCI/CNPQ/MCTIC post-doctoral support. M.Z thanks Fondecyt Regular \#1191505. A.R.A acknowledges support from FONDECYT through grant \#3180203. The authors thank MCTIC/FINEP (CT-INFRA grant 0112052700) and the Embrace Space Weather Program for the computing facilities at INPE. This publication makes use of data products from the Two Micron All Sky Survey, which is a joint project of the University of Massachusetts and the Infrared Processing and Analysis Center/California Institute of Technology, funded by the National Aeronautics and Space Administration and the National Science Foundation. This work has made use of data from the European Space Agency (ESA) mission {\it Gaia} (\url{https://www.cosmos.esa.int/gaia}), processed by the {\it Gaia} Data Processing and Analysis Consortium (DPAC, \url{https://www.cosmos.esa.int/web/gaia/dpac/consortium}). Funding for the DPAC has been provided by national institutions, in particular the institutions participating in the {\it Gaia} Multilateral Agreement. This research has made use of the SIMBAD database, operated at CDS, Strasbourg, France. Some of the plots in this article were prepared using the {\tt python} \citep{CS-R9526} package {\tt matplotlib} \citep{Hunter-2007}.
\end{acknowledgements}

\bibliographystyle{aa}
\bibliography{mylib2020.bib}

\begin{appendix}

\onecolumn
\section{Light curves}\label{sec:appendix}

The VVV-PSF photometry for our sample of Mira and LPV$^{+}$ candidates is provided in Table~\ref{tab:lightcurves}, available at the Centre de Donn\'ees astronomiques de Strasbourg (CDS). A portion is shown here just for guidance in regard to its contents. 

\begin{table*}[!hb]
\footnotesize
\caption{VVV-PSF photometry for our Mira and LPV$^{+}$ candidates\tablefootmark{(a)}\tablefootmark{(b)}}  
\label{tab:lightcurves}      
\centering 
\begin{tabular}{l c c c c r}
\hline
\hline
  \multicolumn{1}{c}{ID} &
  \multicolumn{1}{c}{} & 
  \multicolumn{1}{c}{HJD} & 
  \multicolumn{1}{c}{mag} &
  \multicolumn{1}{c}{$e$(mag)} &
  \multicolumn{1}{c}{bandpass} \\
\hline
308501099316 & * & 58369.027430 & 14.939 & 0.025 & $H$ \\
308501099316 & * & 58373.991807 & 15.000 & 0.027 &  $H$ \\
308501099316 & * & 58340.092523 & 18.090 & 0.166 &  $J$ \\
308501099316 & * & 58369.040062 & 18.896 & 0.173 &  $J$ \\
308501099316 & * & 58374.000068 & 18.293 & 0.165 &  $J$ \\
308503118502 &   & 55497.008828 & 14.529 & 0.070 &  $K_s$ \\
308503118502 &   & 56113.250808 & 14.990 & 0.157 &  $K_s$ \\
\hline
\end{tabular}
\tablefoot{
\tablefoottext{a}{The full table is available at the Centre de Donn\'ees astronomiques de Strasbourg (CDS). A portion is shown here for guidance purposes only.
\tablefoottext{b}{This table contains the following information: column 1 lists the VVV ID, with an asterisk in column 2 indicating the stars that are classified by us as Mira candidates; column 3 gives the Heliocentric Julian Date of the VVV observations; columns 4 and 5 give the corresponding $J$-, $H$-, or $K_s$-band magnitude and associated error, respectively; and the final column indicates the bandpass. We note that measurements with $K_s \lesssim 12$~mag can be considered saturated \citep{ContrerasRamos-2017}, and should thus only be used with extreme caution.}}}
\end{table*}
\FloatBarrier

VVV-PSF $K_s$-band light curves are shown in Figures~\ref{fig:all-light-curves} and \ref{fig:all-light-curves2} for our full sample of Mira candidates and for a representative sample of LPV$^{+}$ candidates, respectively.   

\begin{figure*} [!ht]
\centering
\captionsetup[subfigure]{labelformat=empty}   
\begin{subfigure}{\textwidth} 
\centering
\includegraphics[width=0.61\textwidth]{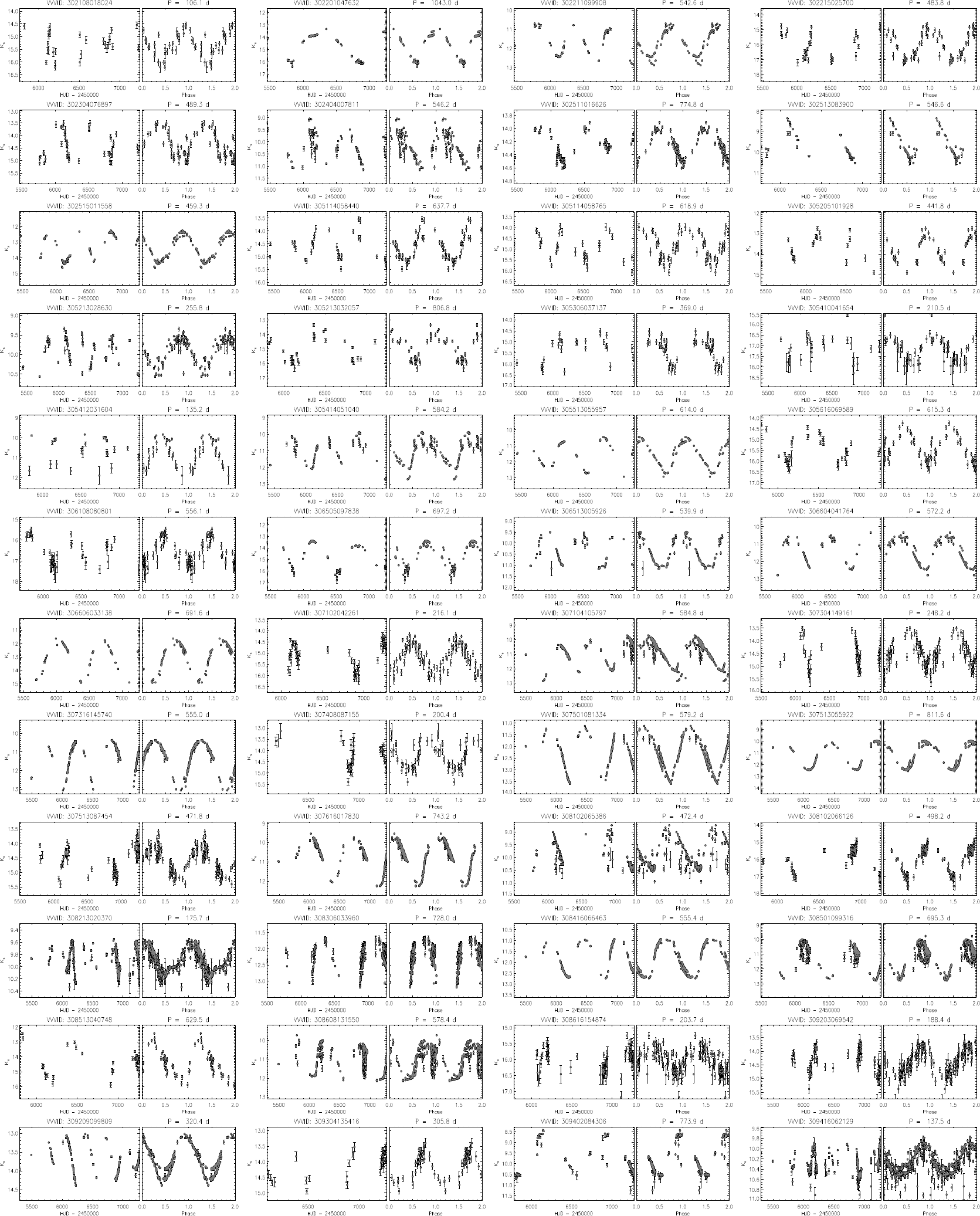}
\end{subfigure}  
   \caption{VVV-PSF light curves in the $K_{s}$ band for the sample of Mira candidates. For each star, both the raw time series (left panels) and phased light curves (right panels) are shown. The VVV ID is displayed on top of the left panel, whereas our derived period is given at the top of the right panel.}
\label{fig:all-light-curves}%
\end{figure*} 

\begin{figure*} 
\centering
\ContinuedFloat
\captionsetup[subfigure]{labelformat=empty}   
\ContinuedFloat
\begin{subfigure}{\textwidth} 
\centering
   \includegraphics[width=\textwidth]{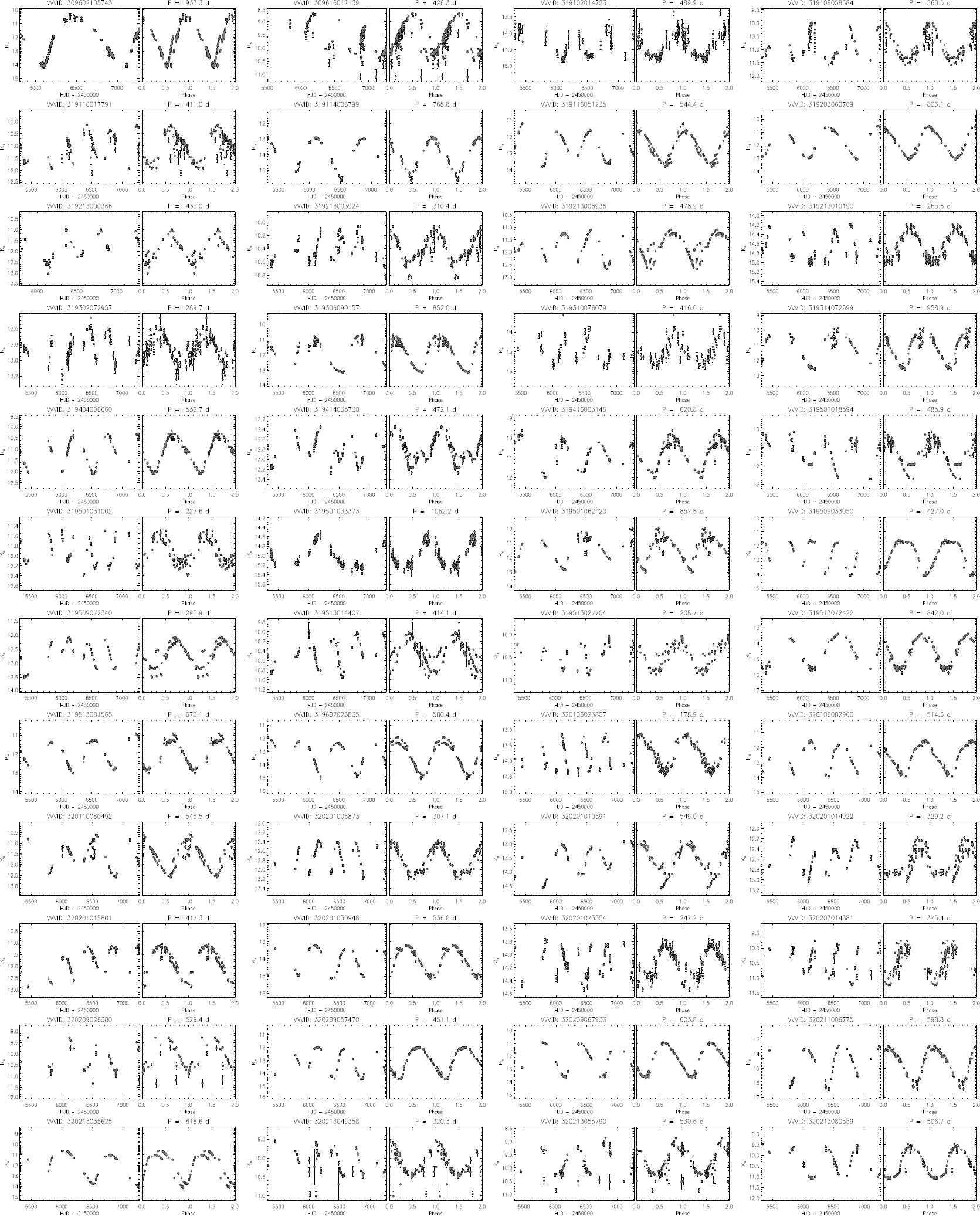}
  \caption{Fig.~A.1. {\em continued}}
\end{subfigure}  
\end{figure*}

\begin{figure*} 
\centering
\ContinuedFloat
\captionsetup[subfigure]{labelformat=empty}   
\ContinuedFloat
\begin{subfigure}{\textwidth} 
\centering
   \includegraphics[width=\textwidth]{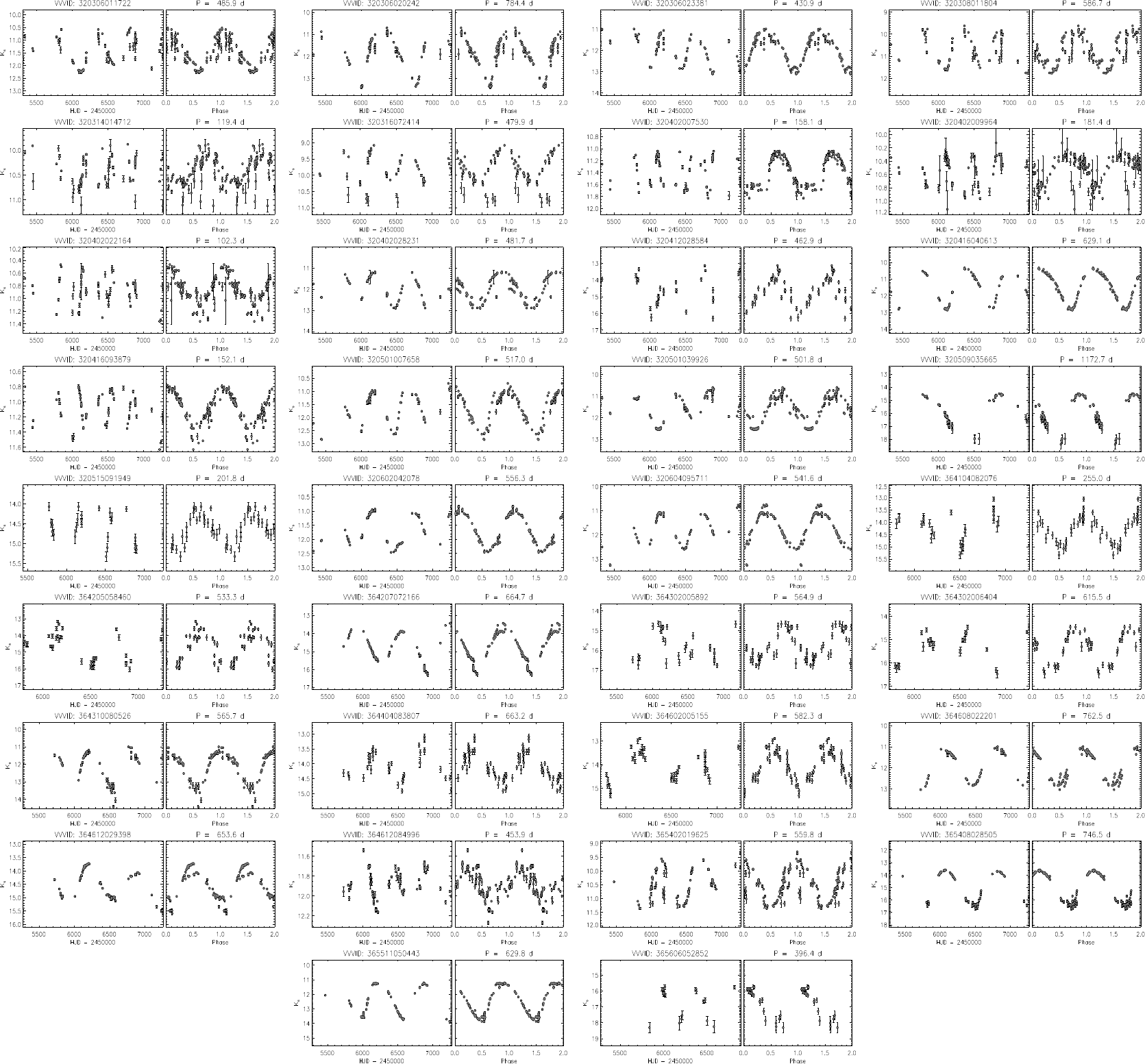}
  \caption{Fig.~A.1. {\em continued}}
\end{subfigure}  
\end{figure*}

\begin{figure*} [!ht]
\centering
\captionsetup[subfigure]{labelformat=empty}   
\begin{subfigure}{\textwidth} 
\centering
\includegraphics[width=\textwidth]{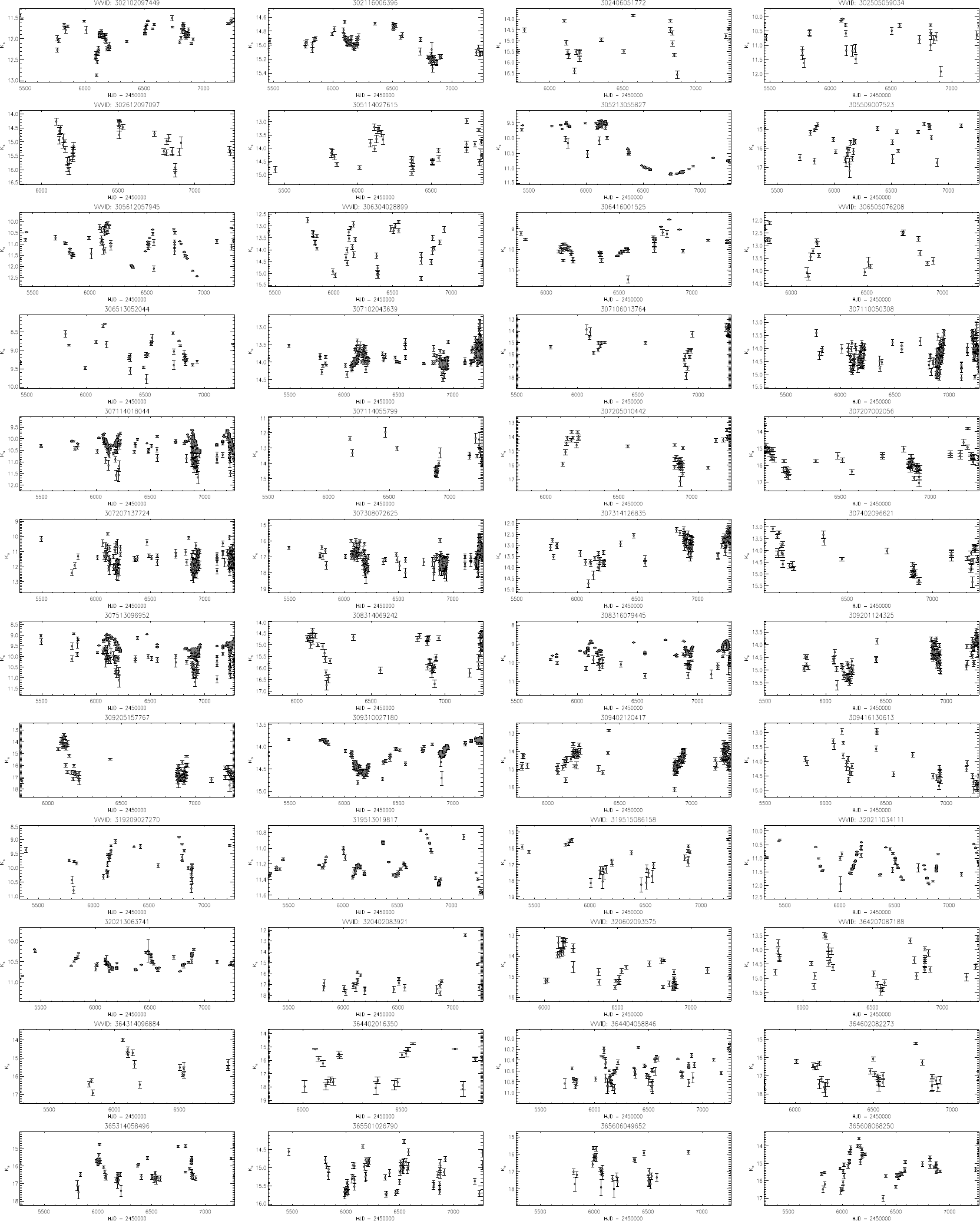}
\end{subfigure}  
   \caption{ Representative VVV-PSF light curves in the $K_{s}$ band for the sample of LPV$^{+}$ candidates. The VVV ID is displayed on top of each panel.}
\label{fig:all-light-curves2}
\end{figure*} 

\FloatBarrier
\newpage
\section{Catalog}\label{sec:catalog}
This appendix reports the main measured photometric parameters for our stars. Columns 1 through 9 in Table~\ref{table:master} list our Mira sources' IDs, coordinates, mean magnitudes, and amplitudes in each of $J$, $H$, and $K_s$, respectively, in addition to their periods, which can be found in column 10. Columns 1 through 8 in Table~\ref{table:master2} provide our LPV$^{+}$ sources' IDs, OGLE IDs, coordinates, mean magnitudes in each of $J$, $H$, and $K_s$, and amplitudes in $K_s$, respectively.

\begin{longtable}{c c c c c c c c c c}
\caption{\label{table:master}      
Catalog of VVV-PSF Mira Candidates$^{(a)}$}\\
\hline\hline
  {VVV ID} &
  {RA} &
  {DEC} &
  {$\langle J \rangle$} &
  {$\Delta{J}$} &
  {$\langle H \rangle$} &
  {$\Delta{H}$} &
  {$\langle K_{\rm s} \rangle$} &
  {$\Delta{K_{s}}$} &
  {$P$} \\
  {} &
  {(J2000)} &
  {(J2000)} &
  {} &
  {} &
  {} &
  {} &
  {} &
  {} &
  {(d)} \\
\hline
\endfirsthead
\caption{{\em continued}.}\\
\hline\hline
  {VVV ID} &
  {RA} &
  {DEC} &
  {$\langle J \rangle$} &
  {$\Delta{J}$} &
  {$\langle H \rangle$} &
  {$\Delta{H}$} &
  {$\langle K_{\rm s} \rangle$} &
  {$\Delta{K_{s}}$} &
  {$P$} \\
  {} &
  {(J2000)} &
  {(J2000)} &
  {} &
  {} &
  {} &
  {} &
  {} &
  {} &
  {(d)} \\
\hline
\endhead
\hline
\endfoot
302108018024     & 265.67709& -34.69817&  &  &  &  & 15.30& 1.31& 106.1\\
302201047632     & 264.60953& -34.51138&  &  &  &  & 14.55& 2.79& 1043.0\\
302211099908     & 265.76817& -34.15634&  &  & 14.26& 2.33& 11.69& 1.74& 542.6\\
302215025700     & 265.84332& -33.79764&  &  &  &  & 16.05& 2.06& 483.8\\
302304076897     & 265.45295& -34.89685&  &  &  &  & 14.22& 1.36& 489.3\\
302404007811     & 265.35269& -34.73811& 15.82& 2.77& 12.63& 2.75& 10.26& 1.52& 546.2\\
302511016626     & 265.73392& -33.90876& 15.60& 0.53& 14.79& 0.53& 14.29& 0.53& 774.8\\
302513083900     & 265.50465& -33.43557& 13.19& 3.06& 10.96& 2.94& 9.34& 1.62& 546.6\\
302515011558     & 265.85518& -33.72611& 17.81& 2.17& 15.36& 1.95& 13.34& 1.94& 459.3\\
305114058440     & 268.35566& -30.01025& 16.07& 1.31& 15.37& 1.31& 14.52& 1.30& 637.7\\	
305114058765     & 268.35643& -30.01044&  &  &  &  & 14.79& 1.86& 618.9\\
305205101928     & 267.66190& -30.46690&  &  &  &  & 14.13& 2.46& 441.8\\
305213028630$^{(b)}$     & 267.88660& -29.87531& 11.87& 1.07& 11.03& 0.73& 10.02& 0.73& 255.8\\
305213032057     & 267.90765& -29.85852&  &  &  &  & 14.82& 2.15& 806.8\\
305306037137     & 267.65753& -30.58666&  &  &  &  & 15.62& 1.69& 369.0\\
305410041654     & 268.06628& -30.00011&  &  &  &  & 17.29& 1.35& 210.5\\
305412031604     & 268.55692& -30.33696& 12.01& 2.01&  &  & 11.03& 2.00& 135.2\\
305414051040     & 268.27928& -29.73206&  &  & 14.41& 4.42& 11.19& 2.43& 584.2\\
305513055957     & 268.12369& -29.64171&  &  &  &  & 11.62& 2.19& 614.0\\
305616069589     & 269.02846& -30.14002&  &  &  &  & 15.36& 1.81& 615.3\\
306108080801     & 269.33742& -29.71888&  &  &  &  & 16.62& 1.55& 556.1\\
306505097838     & 268.58051& -29.09277&  &  &  &  & 14.73& 2.43& 697.2\\
306513005926     & 268.82976& -28.38422& 13.20& 1.98& 11.58& 1.69& 10.42& 1.68& 539.9\\
306604041764     & 269.20789& -29.78146&  &  & 14.22& 3.37& 11.46& 1.97& 572.2\\
306606033138     & 268.80533& -29.24452&  &  &  &  & 13.24& 3.08& 691.6\\
307102042261     & 269.37976& -28.35730&  &  &  &  & 15.33& 1.44& 216.1\\
307104105797     & 269.98035& -28.70129&  &  & 15.15& 3.96& 11.21& 2.18& 584.8\\
307304149161     & 269.88010& -28.59996&  &  &  &  & 14.54& 1.39& 248.2\\
307316145740     & 270.45485& -27.70682&  &  & 15.51& 5.37& 11.71& 2.96& 555.0\\
307408087155     & 270.01501& -28.22376&  &  &  &  & 14.18& 1.55& 200.4\\
307501081334     & 269.18571& -28.02656&  &  & 16.19& 2.56& 12.34& 2.18& 579.2\\
307513055922     & 269.70129& -27.17034&  &  & 14.90& 2.99& 11.21& 2.42& 811.6\\
307513087454     & 269.74841& -27.17961&  &  &  &  & 14.71& 1.07& 471.8\\
307616017830     & 270.57467& -27.55951& 17.70& 3.62& 14.07& 3.60& 10.81& 2.64& 743.2\\
308102065386     & 270.23743& -27.08372&  &  &  &  & 9.78& 1.29& 472.4\\
308102066126     & 270.23846& -27.08387&  &  &  &  & 16.02& 2.29& 498.2\\
308213020370     & 270.35641& -26.04970& 12.26& 0.82& 11.01& 0.78& 9.91& 0.54& 175.7\\
308306033960     & 270.20819& -26.68427& 14.11& 2.55& 13.62& 2.53& 12.71& 2.51& 728.0\\
308416066463     & 271.21903& -26.28482& 17.56& 2.10& 14.38& 1.84& 11.77& 1.83& 555.4\\
308501099316     & 270.00250& -26.81114& 18.87& 5.71& 14.69& 3.37& 11.44& 2.21& 695.3\\
308513040748     & 270.48486& -25.89877&  &  &  &  & 14.38& 3.12& 629.5\\
308608131550     & 271.14381& -26.93477& 14.90& 3.33& 12.50& 3.30& 11.09& 1.91& 578.4\\
308616154874     & 271.50793& -26.41488&  &  &  &  & 16.02& 0.85& 203.7\\
309203069542     & 271.22418& -25.90296&  &  &  &  & 14.50& 1.23& 188.4\\
309209099809     & 271.06545& -25.11297& 17.25& 1.41& 15.10& 1.20& 13.61& 1.19& 320.4\\
309304135416     & 271.44995& -26.08577&  &  &  &  & 14.35& 1.08& 305.8\\
309402084306     & 270.90745& -25.69870& 17.67& 3.38& 12.79& 3.36& 9.61& 1.84& 773.9\\
309416062129     & 272.02021& -24.97945& 12.12& 0.52& 12.04& 0.52& 10.31& 0.51& 137.5\\
309602105743     & 271.19171& -25.66395& 17.92& 3.66& 15.91& 3.64& 11.97& 3.63& 933.3\\
309616012139     & 272.14469& -25.02612& 13.57& 1.44& 11.85& 1.43& 9.67& 1.43& 426.3\\
319102014723     & 266.57016& -30.30780& 18.98& 0.91& 16.37& 0.91& 14.46& 0.90& 489.9\\
319108058684     & 267.35223& -30.39980& 16.57& 1.13& 13.35& 1.13& 11.08& 1.12& 560.5\\
319110017791     & 266.98385& -29.71739& 17.10& 2.52& 13.40& 2.29& 11.24& 1.45& 411.0\\
319114006799     & 267.20926& -29.35417&  &  &  &  & 14.44& 2.54& 768.8\\
319116051235     & 267.82662& -29.68590&  &  & 15.84& 3.02& 12.80& 1.98& 544.4\\	
319203060769     & 266.82190& -30.47402&  &  & 16.00& 3.58& 12.03& 2.48& 806.1\\	
319213000366     & 266.84177& -29.15589&  &  &  &  & 12.00& 1.50& 435.0\\
319213003924     & 266.76257& -29.28409& 19.26& 0.70& 13.69& 0.70& 10.55& 0.51& 310.4\\	
319213006936     & 266.82188& -29.21352&  &  & 16.79& 1.09& 11.90& 1.09& 478.9\\
319213010190     & 266.82281& -29.22846&  &  &  &  & 14.78& 0.69& 265.6\\
319302072957     & 266.42821& -30.40628&  &  & 17.99& 0.38& 13.00& 0.38& 289.7\\
319306090157     & 266.68415& -30.06845&  &  & 15.78& 2.88& 12.02& 2.24& 852.0\\
319310076079     & 266.89575& -29.69910&  &  &  &  & 15.16& 1.66& 416.0\\
319314072599     & 267.15017& -29.33037&  &  & 13.83& 3.01& 11.25& 2.56& 958.9\\
319404006660     & 266.95927& -30.43160& 18.06& 1.72& 14.25& 1.71& 11.33& 1.70& 532.7\\
319414035730     & 267.15457& -29.19770&  &  &  &  & 12.85& 0.78& 472.1\\
319416003146     & 267.67671& -29.37596& 15.81& 1.87& 13.21& 1.86& 11.11& 1.85& 620.8\\
319501018594     & 266.34750& -29.93750& 18.77& 1.86& 14.12& 1.68& 11.43& 1.67& 485.9\\
319501031002     & 266.31407& -30.03889& 19.04& 0.78& 14.47& 0.57& 11.99& 0.56& 227.6\\
319501033373     & 266.36931& -29.96906& 18.95& 0.68& 16.52& 0.67& 15.09& 0.67& 1062.2\\
319501062420     & 266.34757& -30.11633&  &  & 14.68& 3.72& 11.52& 2.27& 857.6\\	
319509033050     & 266.81213& -29.31931& 17.93& 2.40& 14.59& 2.39& 12.66& 2.38& 427.0\\	
319509072340     & 266.87719& -29.37748&  &  & 17.66& 1.72& 12.77& 1.19& 295.9\\
319513014407     & 266.95614& -29.06700& 18.42& 1.68& 13.39& 0.99& 10.55& 0.75& 414.1\\	
319513027704     & 266.99978& -29.07783&  &  & 13.15& 0.92& 10.54& 0.68& 208.7\\
319513072422     & 267.05576& -29.18616&  &  &  &  & 14.59& 2.16& 842.0\\
319513081565     & 267.12584& -29.12312& 17.99& 1.71& 14.83& 1.71& 12.12& 1.70& 678.1\\	
319602026835     & 266.69186& -30.17534&  &  & 17.32& 3.51& 13.42& 2.69& 580.4\\
320106023807     & 267.60669& -28.84640& 18.83& 1.16& 15.89& 1.15& 13.85& 1.15& 178.9\\	
320106082900     & 267.71954& -28.90526&  &  & 16.77& 2.27& 12.75& 2.26& 514.6\\
320110080492     & 267.98277& -28.47871&  &  & 15.48& 2.97& 11.80& 1.75& 545.5\\
320201006873     & 267.00708& -28.95270&  &  & 18.09& 0.80& 12.86& 0.69& 307.1\\
320201010591     & 267.01053& -28.96759&  &  & 18.82& 1.26& 13.62& 1.25& 549.0\\
320201014922     & 267.02805& -28.96702&  &  & 17.88& 0.65& 12.76& 0.65& 329.2\\
320201015801     & 267.05255& -28.93593&  &  & 16.19& 1.56& 11.96& 1.56& 417.3\\
320201030948     & 267.08478& -28.96250&  &  &  &  & 14.15& 1.85& 536.0\\
320201073554     & 267.23621& -28.90871&  &  & 18.72& 0.53& 14.19& 0.52& 247.2\\
320203014381     & 267.54231& -29.25962& 16.05& 1.51& 12.32& 1.46& 10.72& 1.14& 375.4\\
320209026380     & 267.49451& -28.29804& 14.71& 4.17& 11.77& 2.46& 10.28& 1.63& 529.4\\
320209057470     & 267.55177& -28.32888&  &  & 18.12& 3.96& 13.24& 2.52& 451.1\\
320209067933     & 267.55289& -28.36623&  &  & 15.77& 2.70& 12.35& 2.68& 603.8\\
320211006775     & 268.00128& -28.55254& 19.85& 2.70& 18.66& 2.70& 14.94& 2.69& 598.8\\
320213035625     & 267.75684& -27.96231&  &  & 16.62& 3.12& 12.05& 3.12& 818.6\\
320213049358     & 267.76963& -27.99938& 14.71& 0.66& 11.80& 0.66& 10.21& 0.65& 320.3\\
320213055790     & 267.81330& -27.96125& 16.61& 1.36& 12.32& 1.36& 9.79& 1.35& 530.6\\
320213080559     & 267.77519& -28.11835& 16.56& 1.58& 13.04& 1.56& 10.44& 1.43& 506.7\\
320306011722     & 267.42161& -28.70810& 17.73& 1.90& 13.88& 1.88& 11.61& 1.39& 485.9\\
320306020242     & 267.41748& -28.74624&  &  & 16.49& 2.43& 12.01& 2.43& 784.4\\
320306023381     & 267.41463& -28.76234& 20.13& 2.42& 15.47& 2.40& 12.06& 1.76& 430.9\\
320308011804     & 267.98614& -28.95746&  &  & 13.69& 2.16& 10.94& 1.85& 586.7\\
320314014712     & 267.91435& -27.99178& 13.14& 0.58& 11.62& 0.58& 10.62& 0.58& 119.4\\
320316072414     & 268.46198& -28.46651& 12.63& 1.38& 11.38& 1.38& 9.96& 1.37& 479.9\\
320402007530     & 267.28738& -28.90017& 16.36& 0.99& 13.13& 0.86& 11.47& 0.68& 158.1\\
320402009964     & 267.35911& -28.80609& 19.82& 0.56& 14.14& 0.55& 10.56& 0.55& 181.4\\
320402022164     & 267.33350& -28.90544& 15.69& 0.54& 12.50& 0.54& 10.92& 0.54& 102.3\\
320402028231     & 267.41697& -28.81025&  &  & 17.30& 2.51& 12.11& 1.84& 481.7\\
320412028584     & 268.27979& -28.57836&  &  &  &  & 15.00& 2.44& 462.9\\
320416040613     & 268.55517& -28.21039&  &  & 15.07& 4.15& 11.55& 2.45& 629.1\\
320416093879     & 268.67890& -28.22112& 12.97& 0.62& 11.76& 0.62& 11.24& 0.62& 152.1\\
320501007658     & 267.18517& -28.68828&  &  & 15.54& 2.62& 11.74& 1.86& 517.0\\
320501039926     & 267.25619& -28.71666&  &  & 15.43& 3.13& 11.64& 1.73& 501.8\\
320509035665     & 267.61938& -28.14351&  &  &  &  & 16.33& 3.77& 1172.7\\
320515091949     & 268.49466& -28.13295&  &  &  &  & 14.81& 1.01& 201.8\\
320602042078     & 267.59869& -28.94670& 19.91& 2.62& 15.17& 1.81& 11.71& 1.42& 556.3\\
320604095711     & 268.19709& -29.32934& 18.86& 3.82& 14.96& 2.25& 11.92& 1.67& 541.6\\
364104082076     & 266.61044& -25.25956&  &  &  &  & 14.28& 1.64& 255.0\\
364205058460     & 265.85999& -24.49855&  &  &  &  & 14.82& 2.26& 533.3\\
364207072166     & 266.48730& -24.66608& 19.94& 2.77& 18.77& 2.58& 14.73& 2.57& 664.7\\
364302005892     & 265.75815& -24.82447&  &  &  &  & 15.70& 2.19& 564.9\\
364302006404     & 265.75873& -24.82543&  &  &  &  & 15.44& 1.83& 615.5\\
364310080526     & 266.34631& -24.20151&  &  &  &  & 12.24& 2.36& 565.7\\
364404083807     & 266.53552& -25.02552&  &  &  &  & 14.03& 1.28& 663.2\\
364602005155     & 266.02823& -24.77349&  &  &  &  & 14.10& 1.71& 582.3\\
364608022201     & 266.81890& -24.71268&  &  & 15.53& 1.64& 11.93& 1.64& 762.5\\
364612029398     & 266.98517& -24.48362&  &  &  &  & 14.58& 1.44& 653.6\\
364612084996     & 267.13442& -24.45256& 13.96& 1.28& 12.84& 0.76& 11.88& 0.42& 453.9\\
365402019625     & 266.76089& -23.37035& 15.18& 2.88& 12.67& 2.73& 10.48& 1.66& 559.8\\
365408028505     & 267.46152& -23.40129&  &  &  &  & 15.00& 2.73& 746.5\\
365511050443     & 267.52908& -23.01236&  &  & 15.58& 3.81& 12.33& 2.49& 629.8\\	
365606052852     & 267.20936& -23.15345&  &  &  &  & 17.70& 3.71& 396.4\\

\end{longtable}
\tablefoot{
$^{(a)}$ Columns 1 through 10 correspond to their VVV IDs, RA (J2000), DEC (J2000), mean magnitudes and amplitudes in each of $J$, $H$, and $K_s$, and periods, respectively. 
{$^{(b)}$ OGLE-BLG-LPV-075099}}

\vspace{1cm}
\begin{longtable}{ c c c c c c c c}
\caption{\label{table:master2}      
Catalog of VVV-PSF LPV$^{+}$ Candidates$^{(a)(b)}$}\\
\hline\hline
  {VVV ID} &
  {OGLE ID}$^{(c)}$ &
  {RA} &
  {DEC} &
  {$\langle J \rangle$} &
  {$\langle H \rangle$} &
  {$\langle K_{\rm s} \rangle$} &
  {$\Delta{K_{s}}$} \\ 
  {} &
  {} &
  {(J2000)} &
  {(J2000)} &
  {} &
  {} &
  {} &
  {} \\
\hline
\endfirsthead
\caption{{\em continued}.}\\
\hline\hline
  {VVV ID} &
  {OGLE ID} $^{(b)}$ &
  {RA} &
  {DEC} &
  {$\langle J \rangle$} &
  {$\langle H \rangle$} &
  {$\langle K_{\rm s} \rangle$} &
  {$\Delta{K_{s}}$} \\ 
  {} &
  {} &
  {(J2000)} &
  {(J2000)} &
  {} &
  {} &
  {} &
  {} \\
\hline
\endhead
\hline
\endfoot
302404057704&   &265.57148 &-34.65880 &16.74 &16.10& 15.85&  0.82\\
302610077517&  &265.63606 &-33.89896 &  &13.36& 11.54&  1.92\\
305114027615&   &268.35332 &-29.89686 &  &   & 14.17&  1.61\\
305207092343&  &268.16136& -30.77515 &  &  & 15.42&  1.97\\
305304091964&  &268.11481& -31.24289 &  &   & 14.38&  1.88\\
307110120896& 162724 &269.89152 &-27.76557 &12.32 &12.17 &10.53&  1.10\\
307207137724& 163027 &269.89862& -28.16605 &12.23 &   &11.39 & 1.77\\
\end{longtable}
\tablefoot{
$^{(a)}$ The full table is available at the Centre de Donn\'ees astronomiques de Strasbourg (CDS). A portion is shown here for guidance purposes only.
$^{(b)}$ Columns 1 through 8 correspond to their VVV IDs, OGLE IDs, RA (J2000), DEC (J2000), mean magnitudes in each of $J$, $H$, $K_s$, and amplitude in $K_s$, respectively.
$^{(c)}$ When available, OGLE IDs correspond to ``OGLE-BLG-LPV'' followed by the number given. }

\end{appendix}

\end{document}